\pdfoutput=1
\documentclass[a4paper,american,citeautoscript,superscriptaddress,floatfix,pdftex,twoside,%
aps,prb,
 reprint,%
final
]{revtex4-2}%
\usepackage{graphicx}
\usepackage{amsfonts,amsmath,amssymb}
\usepackage{commath}
\usepackage[T1]{fontenc}
\usepackage{graphicx}%
\usepackage{color}
\usepackage{xcolor}
\usepackage[utf8]{inputenc}
\usepackage{microtype}
\usepackage{xspace}
\usepackage{hyperref, hypernat}
\usepackage[none]{hyphenat}
\usepackage{lineno}
\usepackage[displaymath,textmath,graphics]{preview}

\usepackage[version=3]{mhchem}

\graphicspath{{./Fig/}} 
\hypersetup{citebordercolor=yellow,linkbordercolor=red,urlbordercolor=blue} %

\newcommand{\angstrom}{\textup{\AA}}
\newcommand*{\celsius}[1]{\ensuremath{\xspace#1\,^\circ{}\text{C}}\xspace}%
\newcommand*{\degree}[1]{\ensuremath{\ifx\\#1\\\else\xspace#1\,\fi^\circ}\xspace}%
\newcommand{\Toronto}{\affiliation{Departments of Chemistry and Physics, University of Toronto, 80 St. George Street, Toronto M5S 1H6, Canada }}%
\newcommand{\UDE}{\affiliation{Faculty of Physics and Center for Nanointegration Duisburg-Essen (CENIDE), University of Duisburg-Essen, Lotharstr. 1, 47057 Duisburg, Germany}}%
\newcommand{\CAS}{\affiliation{Shanghai Institute of Microsystem and Information Technology, Chinese Academy of Sciences, Shanghai 200050, China}}
\newcommand{\SLAC}{\affiliation{SLAC National Accelerator Laboratory, 2575 Sand Hill Road, Menlo Park, CA, 94025, USA}}

\begin{document}

\title{Femtosecond photo-induced displacive phase transition in Sb$_{2}$Te (group 2) phase-change material}
\author{\mbox{Zhipeng Huang}}
\email[Email:~]{zhipeng.huang@uni-due.de}
\UDE%
\author{Xinxin Cheng}\SLAC%
\author{Hazem Daoud}\Toronto%
\author{\mbox{Wen-Xiong Song}}
\email[Email:~]{songwx@mail.sim.ac.cn}\CAS
\author{\mbox{R. J. Dwayne Miller}}%
\email[Email:~]{dmiller@lphys.chem.utoronto.ca}%
\Toronto%
\author{\mbox{R. Kramer Campen}}
\email[Email:~]{richard.campen@uni-due.de}\UDE

\begin{abstract}\noindent%

Two classes of Phase Change Materials (PCMs) have emerged as the best candidates for applications requiring the fast reading and writing of data: GeTe–Sb$_{2}$Te$_{3}$ pseudobinary alloys (group 1) and doped Sb–Te compounds near the eutectic composition Sb$_{70}$Te$_{30}$ (group 2). 
Both material classes undergo reversible switching between a low-resistance opaque crystalline phase and a high-resistance but less absorbing amorphous phase through heating, electrical, or optical pulses, achieving (sub-)nanosecond switching speeds. 
While group 1 compounds are employed in current generation devices and relatively well studied, model systems in group 2 compounds have been found to crystallize more rapidly and thus offer the perspective of improved devices.  
Despite their superior crystallization speed (SET process), to this point there have been no ultrafast experimental studies on crystallized PCMs of group 2 for the RESET process.
Here we perform ultrafast electron diffraction and femtosecond resolved sum frequency non-linear spectroscopy on Peierls distorted  Sb$_{2}$Te crystallized thin films (PCM of group 2) following femtosecond optical pulse irradiation. 
We observe a pump-induced structural change on two distinct timescales: responses with characteristic timescales of $\approx$ 300 fs and 2~ps. 
We quantified the experimental result by a coherent displacement and the Debye-Waller effect. 
In particular, the $\approx$ 300 fs UED signal results from the ultrafast release of the Peierls distortion through non-thermal coherent Sb displacement, while the 2~ps response reflects electron-lattice equilibrium. 
These results reveal the ultrafast non-thermal structural dynamics of Sb$_{2}$Te and suggest energy-efficient switching of group 2 PCMs should be possible on femtosecond time scales.

\end{abstract}
\maketitle
\section{Introduction}

Phase change materials (PCMs) based on Ge-Sb-Te alloys (\textit{e.g.}\ GeTe, Ge$_{2}$Sb$_{2}$Te$_{5}$, Sb$_{2}$Te$_{3}$, Sb$_{2}$Te, etc.) can be reversibly switched between a low-resistance opaque crystalline phase and a high-resistance but less absorbing amorphous phase. 
As a result they have found wide use as resistive and optical (Blu-ray) nonvolatile memory media ~\cite{Hegedus:intr:2008,Zhang:Intr:2019,Wuttg:Intr:2005,Lankhorst:Intr:2005,Atwood:Intr:2008,Lotnyk:Intr:2019,Kolobov:NC:2011,Nam:science:2012,Loke:Intr:2012,Rao:science:2017}. 
The conductivity switch of PCMs with the most widely used Ge$_{2}$Sb$_{2}$Te$_{5}$ as an example arises because the Fermi level is pinned by defects near the valence band edge in the crystalline phase and near the midgap in the amorphous phase~\cite{Edwards:PRB:2006}. 
Changes in optical properties between the two phases are attributed to differences in bond type~\cite{Fons:PRB:2010,Mukhopadhyay:scirep:2016}.
In particular, the amorphous phase PCMs have a tetrahedrally coordinated metal cation that experiences locally covalent bonding and additionally follows Mott's 8-n rule \cite{Mott:1969,Robertson:bonding:2016}.
In contrast, the crystalline phase PCM metal cations have local octahedral coordination and exhibit resonant bonding (\textit{i.e.,}\ p-type bonding cannot be completely satisfied and thus electronic delocalization and polarizability are increased)~\cite{Shportko:NM:2008,Mukhopadhyay:scirep:2016}.  
To summarize, the resistance of PCMs changes from high to low, and the optical reflectivity in the visible spectrum region changes from low to high during phase transformation from amorphous to crystalline and vice versa~\cite{Shportko:NM:2008,Mukhopadhyay:scirep:2016}.

To act as memory media, the ability to repeatedly record information is required: a SET and RESET process. 
For PCMs writing, \textit{i.e.}\ the SET process, is usually achieved by applying a voltage or optical pulse with sufficient intensity to heat the amorphous film above the crystallization but below the melting temperature.
Information is then \emph{SET} with crystallization of the film. 
Erasing, \textit{i.e.}\ the RESET process, is commonly achieved by the application of a voltage or optical pulse of sufficient energy to melt the PCM film and of sufficiently short duration to allow quenching of the melt.
Such an operation \emph{RESET}s the film in the amorphous state
~\cite{Wuttg:Intr:2005,Lankhorst:Intr:2005}. 

Two classes of PCMs have proven to be the best candidates for practical applications: GeTe–Sb$_{2}$Te$_{3}$ pseudobinary alloys (group 1) and (Ag and/or In doped) Sb–Te compounds near the eutectic composition Sb$_{70}$Te$_{30}$ (group 2)~\cite{Lankhorst:Intr:2005}. 
While group 1 materials are currently employed in devices and their basic physics more widely studied, recent work suggests that the SET operation may be significantly more rapid in  Sb$_{2}$Te, and perhaps group 2 compounds more generally~\cite{Lankhorst:Intr:2005}. 
This potential speed-up is a consequence of different crystallization mechanisms: in group 1 compounds crystalline growth is dominated by nucleation, whereas in \ce{Sb2Te} it occurs via growth from the crystalline edge of the amorphous volume towards the center~\cite{Matsunaga:NM:2011,Shen:Intr:2023}. 
Viewed from the RESET perspective, recent works suggest that crystallized group 1 compounds can be directly amorphized with ultra-short laser pulses, bypassing the melting step~\cite{Kolobov:NC:2011, Hu:ACSNano:2015}. 
Amorphization without melting potentially results in significant energy savings during RESET and, if femtosecond pulses are employed, significant speed up.
To understand whether rapid, low-energy, RESET is possible in group 2 compounds, experiments are required in which the crystalline to amorphous phase transition is probed with femtosecond time and Ångstrom temporal resolution.

Ultrafast electron/X-ray diffraction techniques have been applied to investigate the PCMs of group 1 (\textit{e.g.}\ GeTe,
Ge$_{2}$Sb$_{2}$Te$_{5}$) 
to reveal their atomic structural changes with femtosecond temporal resolution after optical excitation~\cite{Hu:ACSNano:2015,Waldecker:NM:2015,Hada:scirep:2015,Matsubara:PRL:2016,Mitrofanov:scirep:2016,Zalden:science:2019,Qi:PRL:2022}.
Under such conditions a light-induced structural change from rhombohedral to metastable cubic lattice was observed in crystallized GeTe and \ce{Ge2Sb2Te5}~\cite{Hu:ACSNano:2015,Matsubara:PRL:2016}.
Under more intense illumination with femtosecond optical pulses, laser-induced amorphization of crystallized GeTe and \ce{Ge2Sb2Te5} has been observed~\cite{Kolobov:NM:2004,Fons:PRB:2010,Kolobov:NC:2011}.
This non-thermal amorphization has been rationalized by proposing light-excited charge carriers relax into tail states, associated with the weak bonds in the crystal, thus inducing breaking and reorganization of the rhombohedral GeTe solid network~\cite{Kolobov:NC:2011}.
Prior studies have suggested that one possible \emph{mechanism} of amorphization involves an initial transition to a metastable cubic phase implying that the reversible, light-induced, rhombahedral$\leftrightarrow$cubic transition in group 1 compounds is relevant for ultrafast amorphization~\cite{Hu:ACSNano:2015}.

Moreover, theoretical and experimental studies suggested that the lattice transforms from rhombohedral to cubic were different between GeTe and \ce{Ge2Sb2Te5} in group 1 compounds.
In GeTe the A$_{1g}$ phonon mode has been argued to be strongly coupled to the electronic excitation and responsible for the initial ultrafast local lattice change~\cite{Matsubara:PRL:2016, Chen:PRL:2018, Chen:JPCL:2023}.
Following the initial local change in symmetry, a global structural change to cubic is observed~\cite{Hu:ACSNano:2015,Matsubara:PRL:2016}. 
The rock-salt \ce{Ge2Sb2Te5} has local Peierls distortions and 20\% vacancy concentrations.
Here only a local structural transition from the rhombohedral to the cubic geometry happening within 0.3~ps through atomic motion along the $\langle111\rangle$ direction was observed.
\ce{Ge2Sb2Te5} has random local distortions along the degenerate $\langle111\rangle$ direction.
As a result, global symmetry changes are not possible and the A$_{1g}$ mode responsible for the ultrafast structural change does not exist. 
Initial studies on the ultrafast structural dynamics of crystalline \ce{Ge2Sb2Te5} concluded that the mechanism of laser-induced structural change was purely thermal as all the diffraction peak intensities decreased on a time scale of a couple of ps and can be described well by considering only the Debye-Waller effect~\cite{Waldecker:NM:2015, Hada:scirep:2015}.
More recent results, from some of the same investigators, have concluded that \ce{Ge2Sb2Te5} undergoes a local \emph{non-thermal} structural transition from rhombohedral to cubic~\cite{Qi:PRL:2022}, and that this change is driven by a highly damped, Raman-inactive, phonon~\cite{Qi:PRL:2022, Miller:PRB:2016}.
Regardless of the mechanism, studies on both GeTe and Ge$_{2}$Sb$_{2}$Te$_{5}$ have concluded that light-induced structural changes are initiated by the movement of Ge atoms~\cite{Matsubara:PRL:2016, Qi:PRL:2022, Hada:scirep:2015}.
However, both raffling motion~\cite{Matsubara:PRL:2016} and displacive motion~\cite{Qi:PRL:2022,Chen:PRL:2018} of Ge atoms have been proposed.

While in group 1 compounds the existence of laser-induced, non-thermal, structural change seems clear (even if the mechanism by which this motion proceeds remains controversial) the situation for group 2 compounds is substantially less clear. 
As noted above in group 1 compounds structural change proceeds by the movement of Ge atoms. 
The presence of Ge (relative to Sb in group 2 compounds) is expected to create a larger local Peierls distortion (in the unpumped sample) and thus a larger driving force for structural change in group 1 than group 2 compounds~\cite{Pieterson:JAP:2005, Shen:Intr:2023}.
Because group 2 materials offer the possibility of faster SET operations than group 1, \textit{i.e.}\ faster crystallization, resolving whether ultrafast, non-thermal (and thus low energy) RESET is possible for these materials is important. 

As discussed above it is clear that inducing ultrafast, displacive motion in group 2 PCMs requires changes in both electronic and lattice structure and structural symmetry. 
Here, we investigate the ultrafast structural and carrier dynamics of crystallized Sb$_{2}$Te by ultrafast electron diffraction (which probes the ultrafast structural dynamics with Ångstrom spatial and femtosecond temporal resolution)~\cite{Siwick:science:2003, Zong:MRSreview:2021} and pump-probe sum frequency generation spectroscopy (which probes charge carrier and structural symmetry dynamics with femtosecond temporal resolution)~\cite{Shen:SFG:1989, Foglia:APL:2016}.
In the UED experiments we observe an ultrafast non-thermal solid-solid phase transition after excitation by 343 nm femtosecond laser pulses.
The diffraction peak intensities decrease anisotropically with a time constant of hundreds of femtoseconds after optical excitation. The diffraction peaks -- \textit{e.g.}\ $(111)$, $(311)$ and $(511)$ -- that originate mainly from the Peierls distortion, show large decreases in amplitude in approximately 300 fs.
Careful structure factor calculations clarify that the observed signal can be rationalized as a coherent displacement of Sb atoms along the [111] direction that suppresses the Peierls distortion and causes the local rhombohedral to the cubic structure change.
After several picoseconds diffraction peak intensities decrease further and became nearly isotropic as the Debye-Waller effect becomes dominant, suggesting that electron-phonon equilibrium has been achieved.
Combining the structure factor change due to the Sb atomic coherent displacement along [111] direction and the Debye-Waller effect, we quantitatively determined the Sb atomic coherent displacement amplitude and mean-square displacement as a function of pump-probe delay by fitting the calculated results with the experimental diffraction intensity change. 
We demonstrate that the coherent atomic displacement and mean square displacement amplitudes change linearly as a function of pump laser fluence and that the initial rhombahedral to cubic transformation is likely the first step in amorphization.
Time-resolved SFG experiment on the crystallized Sb$_{2}$Te thin films show that carrier cooling occurs on the same time scale as electron-phonon equilibrium and confirm that the UED structure evolution describes a change to higher symmetry.
Taken as a whole these results indicate that \ce{Sb2Te}, and by extension all group 2 PCMs, are good candidates for an ultrafast, low-energy RESET operation in memory applications.

\section{Experimental Methods}
\subsection{Sample Preparation}
The 50~nm thick Sb$_{2}$Te for UED experiments was prepared by RF sputtering~\cite{Wang:Sb2Te:2018,Zhu:Sb2Te:2019}. 
The Sb$_{2}$Te thin film was deposited on an ultra-thin (3~nm) carbon film supported by a 400 mesh copper grid.
The as-deposited film is amorphous.
Following deposition, we anneal the sample at a temperature of 150\celsius~ for 10 minutes. 
The amorphous sample crystallized to a distorted cubic crystalline structure as confirmed by the static electron diffraction shown in~\autoref{fig:setup}c.
The 50~nm thick Sb$_{2}$Te thin film on a SiO$_{2}$ substrate for pump-probe SFG experiments is also prepared by RF sputtering and then annealed at 150\celsius~ for 10 minutes. 

\subsection{Femtosecond Electron Diffraction Setup}
A Yb:KGW femtosecond laser operating at a 1 kHz repetition rate with a center wavelength of 1030 nm and a 160 fs (FWHM) pulse duration was used as the master laser. 
The laser beam is then split into two, as shown in~\autoref{fig:setup}(a). 
One of the beams is frequency upconverted to the fourth harmonic using two BBO crystals, achieving a center wavelength of 257 nm. 
This beam is then directed to the femtosecond electron gun chamber to illuminate the back of a photocathode (30 nm Au thin film) to generate femtosecond electron pulses, which are accelerated by a DC electric field with a voltage of 100 kV. 
The accelerated 100 keV electron pulses are then passed through a pinhole of the grounded anode plate, which isolates the electron source chamber from the sample chamber for differential pumping to achieve a vacuum pressure below 10$^{-7}$ mbar for the electron gun to avoid electrical discharge at the applied high potential of 100 kV.
The sample is mounted on a three-axis translational stage, with a sharp knife edge at the top of the sample holder for measuring the optical pump and electron probe beam sizes and checking the spatial overlap of these two beams.
The second laser beam is frequency tripled to a center wavelength of 343 nm and sent to the sample chamber for pumping the sample. 
The transmitted scattered electrons from the samples are collimated by a solenoid magnetic lens and detected by a phosphor screen and camera. 
The delay between the optical pump pulse and the electron probe pulse is precisely controlled by a delay stage. 
The pump pulse energy is controlled by rotating the half-wave plate placed before the polarizer.
The electron source and sample chambers are evacuated by a turbomolecular pump backed by a scroll pump.
The two-stage pumping achieves a typical vacuum condition of 10$^{-8}$ mbar in the electron source chamber and 10$^{-7}$ mbar in the sample chamber.

The detected electron diffraction intensity $I_{hkl}$ of a crystallized sample is proportional to the square of the structure factor $F_{hkl}$~\cite{Coppens:ITC:2006}. 
\begin{equation}
	\label{eq:diffI}
	I_{hkl} \propto |F_{hkl}|^2
\end{equation}
where $F_{hkl}$ is the sum of the amplitude function from each atom $j$ in the unit cell.
\begin{equation}
	\label{eq:strucI}
	F_{hkl}=\sum_{j=1}^{N}f_j\exp[2\pi\cdot i(hx_j+ky_j+lz_j)] 
\end{equation}
where $f_j$ is the atomic scattering factor of atom $j$ in the unit cell. $x_j$, $y_j$, $z_j$ are coordinates of the atom $j$ in the unit cell. 
If there is any change in the position of the atom in the unit cell induced by the pump beam, the diffraction intensity will change accordingly. 

\subsection{Femtosecond-resolved Sum Frequency Generation Spectroscopy}
We employ a laser system composed of a Ti: Sapphire oscillator and a regenerative amplifier for the time-resolved SFG spectroscopy measurement.
The amplifier delivers a beam with an energy of $6\frac{\text{mJ}}{\text{pulse}}$, a center wavelength of 800 nm, a duration of 35~fs (FWHM), and a repetition rate of 1 kHz. 
It is further split into three beams. 
One beam with an energy of $1.8\frac{\text{mJ}}{\text{pulse}}$ pumps an optical parametric amplifier (OPA). 
The produced signal and idler beams from the OPA are subsequently mixed in a collinear difference frequency generation scheme to generate broadband infrared (IR) femtosecond pulses centered at 5.5 \textmu m.

The second beam is spectrally shaped by an air-spaced etalon to produce a narrowband 800 nm beam and the third used as the pump pulse in this study. 
The pulse energies used for the 800 SFG and 800 pump are $1.12\frac{\mathrm{\text\textmu J}}{\text{pulse}}$ and $2.05\frac{\mathrm{\text\textmu J}}{\text{pulse}}$, respectively.  
To conduct time-resolved SFG measurements, the infrared and narrow-band 800 nm beams are spatially and temporally overlapped on the sample surface as illustrated in \autoref{fig:fastdyn}(d). 
The time delay between the pump pulse and the probe pulse pair is then scanned using a motorized delay stage in the optical path of the pump beam. 
The generated SFG beam is guided to a spectrograph to disperse the beam onto an emICCD camera. 
A detailed description of the optical beam path can be found in our recent paper~\cite{Huang:RSI:2024}. 

We choose 5500~nm for the IR wavelength in this experiment because this photon energy is smaller than the bandgap of the crystallized Sb$_{2}$Te thin film and higher than any phonon frequency. 
The resulting SFG signal is thus non-resonant, \textit{i.e.}\ responds instantaneously to \emph{all} structural and electronic change without convoluting relaxation of particular states. 
The pump beams used for the UED experiments and the femtosecond-resolved SFG experiments are both higher than the bandgap of crystallized Sb$_{2}$Te sample. They both induce interband transitions and generate free charge carriers. 
Thus the dynamics induced by both pump beams should be quite similar. 
The polarization of all beams is set to P polarized. 

The SFG intensity is proportional to the second-order susceptibility, \textit{i.e.} the $\chi^{(2)}$, of materials.
In the dipole approximation $\chi^{(2)}$ is identically zero~\cite{Shen:SFG:1989} in solids with inversion symmetry. 
The macroscopic $\chi^{(2)}$ is proportional to the microscopic first hyperpolarizability $\beta$.    
\begin{equation}
	\label{eq:hyperpo}
	\beta \propto \frac{\partial \alpha_{ij}}{\partial Q_q}\cdot\frac{\partial \mu_k}{\partial Q_q}
\end{equation}
where $\frac{\partial \alpha_{ij}}{\partial Q_q}$ is the Raman polarizability and $\frac{\partial \mu_k}{\partial Q_q}$ is the IR transition dipole. 
If the pump beam modifies either term it will change the SFG intensity.
The macroscopic SFG signal is proportional to the orientationally averaged $\beta$. 
If the pump beam also modifies the orientation (of the unit cell), it will also change the SFG intensity.

\section{Results and Discussion}
\begin{figure*}[!htbp]
	\centering
	\includegraphics[width=0.85\textwidth]{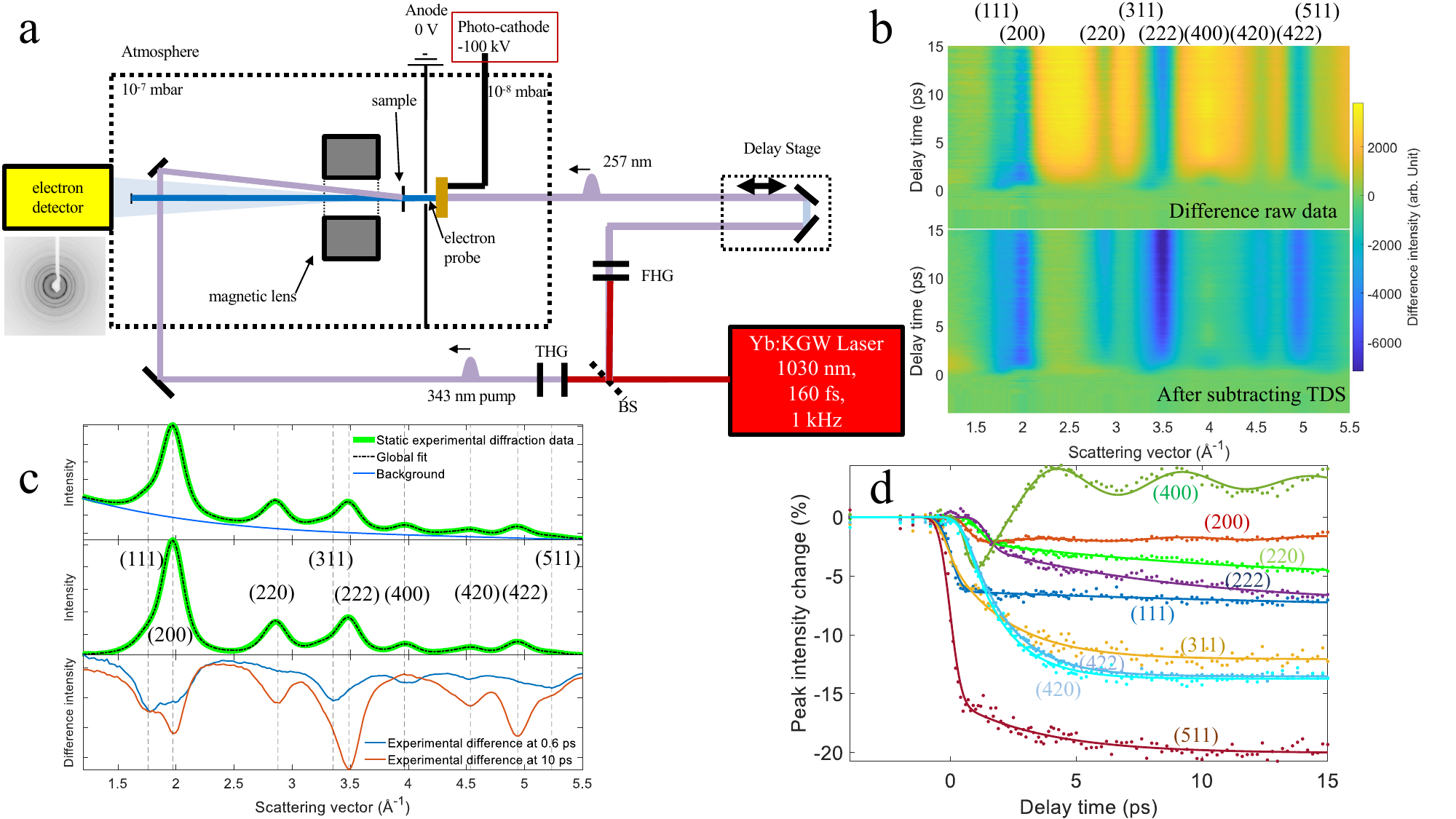}
	\caption{(a) Schematic drawing of the experimental setup for ultrafast electron diffraction
	on Sb$_{2}$Te. (b) The radial average of the Sb$_{2}$Te ultrafast electron diffraction experimental raw difference
 results $I_{\text{diff}}$ (top) and the results after subtracting thermal diffuse scattering (TDS) background (bottom) as a function of pump-probe delay time.
 ($I_{\text{diff}} = I_{\text{pump}}-I_{\text{probe}}$, where $I_{\text{pump}}$ is the total scattering intensity of Sb$_{2}$Te under pump pulse excitation,
 $I_\text{probe}$ is the total scattering intensity of Sb$_{2}$Te without pump pulse excitation). The pump fluence was set to 1.5 mJ/cm$^2$.
	(c) The radially averaged Sb$_{2}$Te static electron diffraction result and its global fitted result for subtracting the background (top panel). 
	The background-subtracted radially averaged Sb$_{2}$Te static electron diffraction and indexing of observed peaks (middle panel). 
	The thermal diffuse scattering background subtracted experimental difference results with UV pump pulse 0.6~ps and 10~ps earlier than femtosecond electron probe pulse (bottom panel). 
	(d) Bragg peak relative intensities as a function of pump-probe delay time from
-4 ps to 15 ps. See texts for details.}
	\label{fig:setup}
\end{figure*}

The static Debye–Scherrer electron powder diffraction pattern of Sb$_{2}$Te without pump pulse excitation is shown in~\autoref{fig:setup}(a). 
Sharp diffraction rings are clearly observed, consistent with the expected long coherence length of our femtosecond electron beam relative to the sample thickness~\cite{Ishikawa:science:2015}. 
\autoref{fig:setup}(c) top panel shows the measured radially averaged static electron diffraction pattern and the result of a globally fitted background to all diffraction peaks used to subtract the inelastic and atomic scattering background~\cite{Zahn:SD:2021}. 
\autoref{fig:setup}(c) middle panel shows the background subtracted radially averaged Sb$_{2}$Te static electron diffraction pattern. 
Sb-Te chalcogenide phase change materials annealed at temperatures equal or lower than 150\celsius~ typically result a Peierls distorted cubic structure~\cite{Zheng:nanores:2016}. 
As discussed in detail in the Supporting Information, relatively intense (111), (311) and (511) peaks are diagnostic of this structure. 
The experimentally observed diffraction peaks, as labeled in \autoref{fig:setup}(c) middle panel, are consistent with the expected structure.

The bottom panel of \autoref{fig:setup}(c) shows the experimental difference results at two pump-probe delays with the thermal diffuse scattering background subtracted. 
The blue and red curves correspond to UV pump pulses arriving 0.6 ps (blue) and 10 ps (red) earlier than the femtosecond electron probe pulse. 
Clearly, the (111), (311), and (511) peaks show a large decrease at 0.6~ps, although their static diffraction peak intensities are very small. 
If the decrease of these peak intensities is a thermal effect, their dynamics are expected to follow the Debye-Waller relationship~\cite{Peng:DW:2004}:
\begin{equation}
	\label{eq:dwe}
	\frac{I_{hkl}(t)}{I_{hkl}(\text{probe})} = \text{exp}\left[-\frac{q^{2}}{3}\left(\langle u^{2}\rangle_{(t)}-\langle u^{2}\rangle_{(\text{probe})}\right)\right]
\end{equation}
where $I_{hkl}$ is the intensity of the $(hkl)$ diffraction peak, $q$ is the scattering vector and $\langle u^{2}\rangle$ is the atomic mean square displacement (MSD).  
According to \autoref{eq:dwe}, for peaks with similar $q$, the stronger the static diffraction peak, the larger the experimental difference intensity ($I_{hkl}(t)-I_{hkl}(\text{probe})$) will be. 
The (200), (222), and (422) static diffraction peaks are much stronger than their neighbors: the (111), (311), and (511) respectively. 
If the pump-induced change in diffraction peak intensities is only influenced by the Debye-Waller effect, the pump-induced intensity change of peaks (200), (222), and (422) should be larger than those of peaks (111), (311), and (511).
Clearly this is not the case at 0.6~ps delay: peaks (111), (311), and (511) show by far the largest change.
Thus, apparently, at least at sub-ps timescales \ce{Sb2Te} undergoes non-thermal structural change following optical excitation. 
After some delay we expect the electronic and phononic system of \ce{Sb2Te} to fully thermalize and, therefore, the Debye-Waller effect to become dominant in the pump-induced diffraction data.
In such a scenario the pump-induced change in diffraction peak intensity should be larger for (200), (222), and (422) than for their neighboring (111), (311), and (511).
Reference to \autoref{fig:setup}(c) clearly shows this condition has been achieved as of a 10 ps delay.

While the qualitative analysis shows that laser-induced non-thermal structural occurs, it does not quantify the structural change and thus does not reveal the particular atomic motions that are induced. 
The first step in a quantitative analysis that meets these challenges is to remove the thermal diffuse contribution to the raw difference diffraction signal.
\autoref{fig:setup}(b) top panel shows the radially averaged experimental difference raw data, obtained by subtracting the probe-only data from the radially averaged pump-probe diffraction result to highlight the influence of pump pulse excitation. 
In addition to the decrease in all diffraction peaks, the regions between the diffraction peaks increase due to thermal diffuse scattering~\cite{Stern:PRB:2018, Durr:MRS:2021}. 
\autoref{fig:setup}(b) bottom panel shows the radially averaged experimental difference diffraction pattern after subtracting the thermal diffuse scattering background. 
The procedure for subtracting the thermal diffuse scattering (TDS) background involves first globally fitting all diffraction peaks~\cite{Zahn:SD:2021}, as shown in \autoref{fig:setup}(c), to precisely subtract the background. 
Then, the background-removed probe-only data is subtracted from the background-removed pump-probe diffraction results.
\autoref{fig:setup}(d) shows how the observed diffraction peak intensities change as a function of the pump-probe delay time.
The experimental results are fitted with exponential functions convoluted with instrument Gaussian functions to derive the time constants of the evolution of different diffraction peaks. 
The solid curves show the fitted results. 
The equation used to fit the diffraction peak intensities as a function of the pump–probe delay time is explained in the supplementary information. 
Two qualitative features are clearly observed: (1) the intensities of peaks (111), (311), and (511) clearly evolve on two characteristic time scales; (2) The Sb$_{2}$Te sample reaches a quasi-equilibrium state at $\approx$ 5 ps after the femtosecond pulse excitation. 
These observed results are consequences of both structural change and thermal evolution, which will be quantitatively modeled and discussed below.
\begin{figure*}[!htbp]
	\centering
	\includegraphics[width=0.85\textwidth]{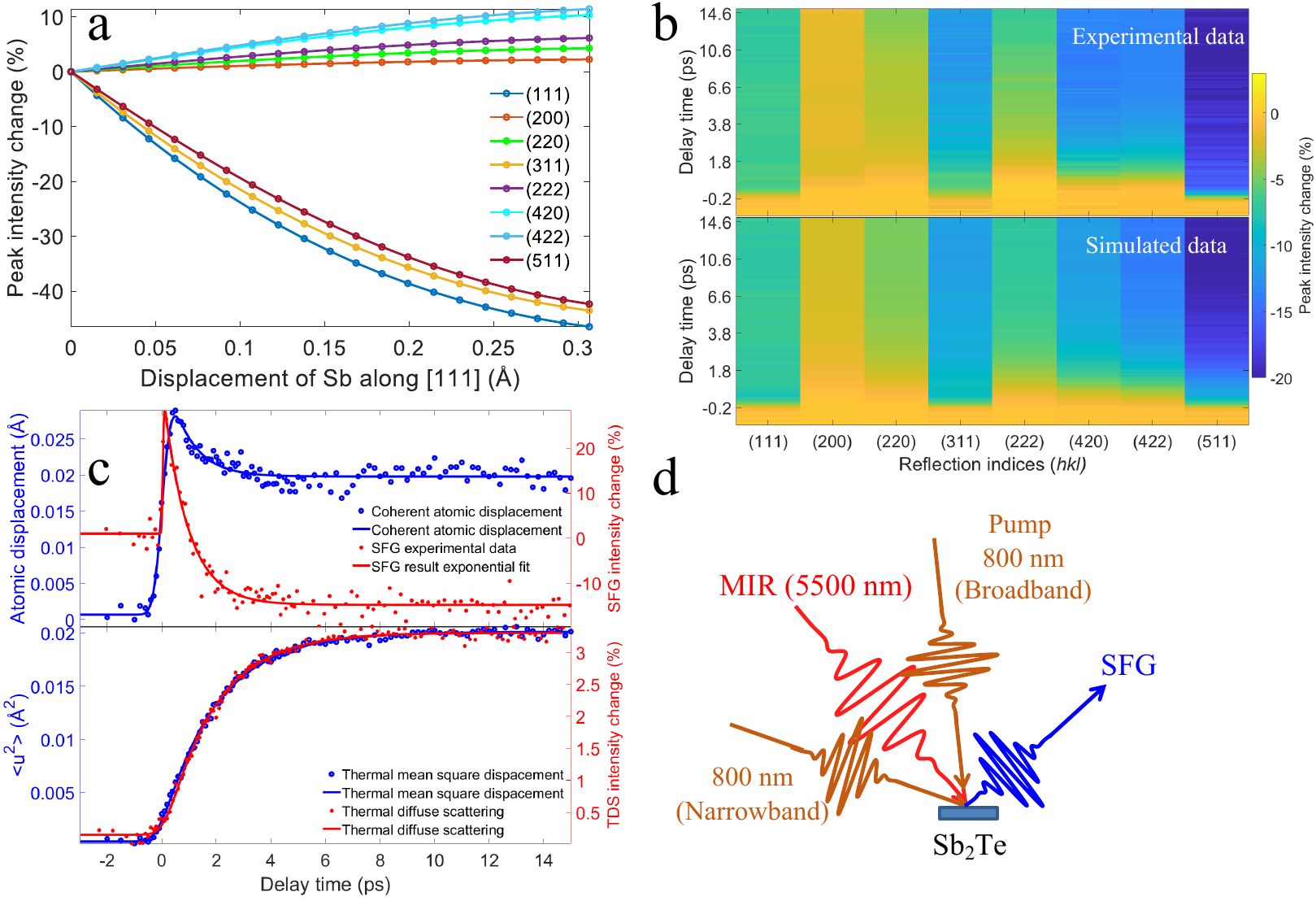}
	\caption{(a) The structure factor calculated relative intensity change of observed Bragg reflections
 by displacing the Sb (which is located in the center of the distorted face-centered cubic lattice unit cell) along the [111]
 direction (to a less distorted cubic lattice symmetry).
	(b) Comparison of the experimental and calculated results of the Bragg peak relative intensities change as 
 a function of pump-probe delay time from -4~ps to 15~ps.
    (c) The derived coherent atomic displacement and pump-probe sum frequency generation spectrum intensity
    as a function of pump-probe delay time (top). The derived atomic mean square displacement $\langle u^{2}\rangle$ and thermal
    diffuse scattering relative intensity as a function of pump-probe delay time (bottom).
    (d) The scheme of the pump-probe sum frequency generation experiments.
	See texts for details.}
	\label{fig:fastdyn}
\end{figure*}

We first address the non-thermal effect responsible for the ultrafast dominant decrease of peaks (111), (311), and (511) to connect the observed changes to a quantitative structural picture. 
We start with the hypothesis that non-thermal structural change causes the Peierls distorted cubic structure to move to a higher symmetry as suggested by the results from the PCMs of group 1 (\textit{e.g.} GeTe~\cite{Hu:ACSNano:2015} and \ce{Ge2Sb2Te5}~\cite{Qi:PRL:2022}). 
Since the Peierls distortion in PCMs exists along the [111] direction~\cite{Gaspard:Peierls:2016}, we conducted structure factor calculations by displacing the distorted atom along the [111] direction (and thus releasing the Peierls distortion in the unit cell and evolving the distorted cubic unit cell towards metastable cubic).
The procedure for calculating the structure factor is shown in the Supporting Information. 
\autoref{fig:fastdyn} (a) shows the diffraction peak intensity changes under such a displacement of the distorted Sb atom. 
Clearly peaks (111), (311), and (511) decrease in intensity, while other peaks show little change or slightly increase when the Sb atom is displaced along the [111] direction.
The observed pattern does \emph{not} quantitatively match the peak intensity changes shown in \autoref{fig:setup}(d).
One possible explanation for this discrepancy is that phonon thermalization occurs concurrently with non-displacive excitation: \emph{both} Sb coherent displacement \emph{and} the Debye-Waller effect should be accounted for in the structure factor calculation. 
\autoref{fig:fastdyn}(b) bottom panel shows the calculated results accounting for both effects.
Clearly doing so quantitatively reproduces the experimental data (top panel). 

\autoref{fig:fastdyn}(c) bottom panel compares the derived mean square displacement and the thermal diffuse scattering (integration of the region between peaks (200) and (220) of the difference raw data as shown in \autoref{fig:setup}(b) upper panel). 
They show the same rising onset and single exponential rising time constant, which confirms that both the Debye-Waller effect and the thermal diffuse scattering originate from lattice heating by electron-phonon scattering.
As noted in the Introduction, laser excitation has been argued to produce a rattling motion~\cite{Matsubara:PRL:2016} of Ge atoms in  Ge$_{2}$Sb$_{2}$Te$_{5}$ and GeTe.
If the laser-induced motion of Sb atoms resulted in a rattling motion, one would expect the diffuse scattering background to \emph{increase} when the rattling motion starts (rattling leads to disorder~\cite{Wall:VO2:2018,De:NP:2023}). 
However, our observed thermal diffuse background rises only thermally, \textit{i.e.}\ on ps timescales. 
Our data thus clearly suggest a coherent atomic displacement, and not a rattling of Sb, explains Sb motion in Sb$_2$Te on 300 fs timescales.

While the time-resolved diffraction measurements allow access to structure change, they do not probe charge carrier dynamics and only indirectly probe structural symmetry.
We thus next utilized pump-probe sum frequency generation spectroscopy to further characterize the carrier dynamics and structural symmetry change after laser excitation of a 50~nm polycrystalline Sb$_{2}$Te thin film deposited on a SiO$_{2}$ substrate~\cite{Shen:SFG:1989,Foglia:APL:2016}. 
The scheme of the pump-probe sum-frequency generation spectroscopy experiment is shown in \autoref{fig:fastdyn}(d)~\cite{Huang:RSI:2024}: we excited the Sb$_{2}$Te thin film with a 35~fs laser pulse centered at 800~nm and probed the charge carrier and structural symmetry dynamics using a MIR femtosecond beam (centered at 5500~nm) and a narrowband 800~nm beam. 

\begin{figure*}[!htbp]
	\centering
	\includegraphics[width=0.78\textwidth]{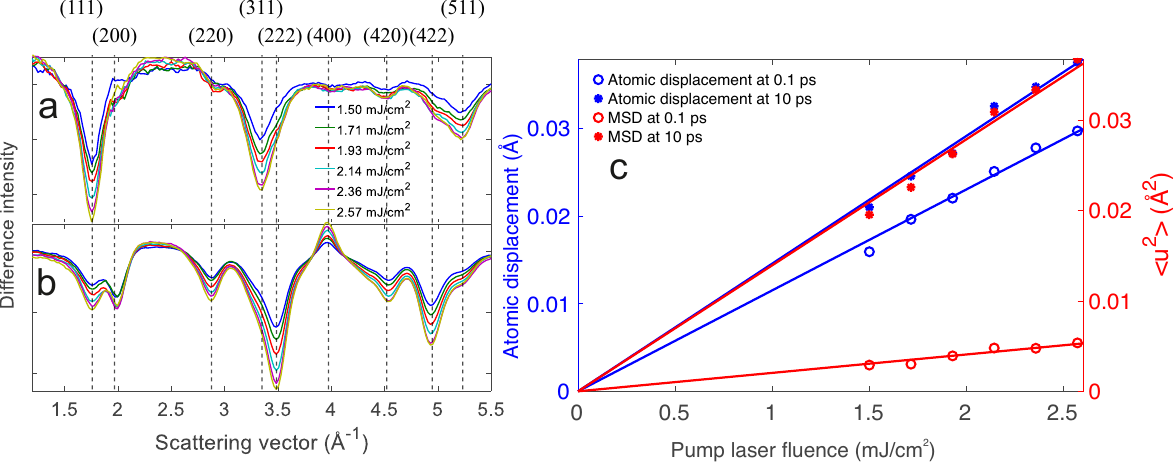}
	\caption{(a) The radially averaged experimental difference diffraction data under different pump fluence excitation
 with a fixed pump-probe delay time at 0.1~ps. (b) The radially averaged experimental difference data under
 different pump fluence excitation with a fixed pump-probe delay time at 10~ps. (c) The derived coherent atomic
 displacement amplitude (blue) and atomic mean square displacement (red) at a fixed delay time of 0.1~ps and 10~ps as a function of pump laser fluence. 
	See texts for details.}
	\label{fig:fluence}
\end{figure*}
The right axis of \autoref{fig:fastdyn}(c) top panel shows the time-resolved SFG result.
Both the $\Delta \text{I}_{\text{SFG}}$ and the derived atomic coherent displacement increase on a timescale of hundreds of femtoseconds and decay within a couple of picoseconds. 
Since pump-probe SFG detects both the charge carrier and structural dynamics, the ultrafast increase of SFG intensity in the femtosecond time scale is mainly due to the sudden increase in the density of charge carriers after laser excitation~\cite{Hohlfeld:CP:2000}. 
Above band gap excitation of \ce{Sb2Te} is expected to result in a transient increase in free carrier population, and thus an increase in polarizability and $\text{I}_{\text{SFG}}$ as suggested in equation \ref{eq:hyperpo}.
As inspection of \autoref{fig:fastdyn}(c) makes clear, our results are consistent with this expectation.
As discussed above in the context of the UED results and SFG signal, the picosecond decay in $\Delta \text{I}_{\text{SFG}}$ is expected to result both from charge carrier relaxation and crystals structural symmetry change. 
The atomic coherent displacement increases the symmetry of the crystal from a distorted cubic to the direction of a metastable cubic.
The sample changes to an inversion symmetric lattice and thus reduces the $\chi^{(2)}$ due to the cancellation of contribution from bulk, which is the reason that the SFG intensity at the quasi-equilibrium state is lower than the static SFG intensity.

While both the UED (and SFG) experiments clearly indicate pump-induced non-thermal atomic coherent displacement and the Debye-Waller effect, they neither offer a clear connection to amorphization nor clarify whether the observed dynamics result from carrier-carrier interaction.
To address both of these questions, we next explore how the amplitudes of the atomic coherent displacement and the mean square displacement change as a function of pump laser fluence.  
\autoref{fig:fluence}(a) and (b) show the thermal diffuse scattering subtracted radially averaged experimental difference diffraction results under different pump fluences with a fixed pump-probe delay time at 0.1~ps and 10~ps, respectively. 
As discussed above, the coherent Sb atomic displacement along [111] happens on the femtosecond time scales and reduces the intensity of peaks (111), (311), and (511) significantly (see \autoref{fig:fastdyn}(a)). 
Inspection of \autoref{fig:fluence}(a) clearly shows that the size of this decrease at 0.1~ps delay increases linearly with increasing pump laser fluence. 
At 10~ps, the lattices have mostly equilibrated with the excited electrons. 
As shown in \autoref{fig:fastdyn}(b) and discussed above, the (200), (222) and (422) peaks are dominant in static diffraction and, thus, decrease more than their neighboring peaks due to the Debye-Waller effect. 
Reference to \autoref{fig:fluence}(c) makes clear that also for these features, which dominate the observed signal at 10~ps delay, the observed intensity change in diffraction features depends linearly on pump fluence.
The linear relationship with pump laser fluence at both timescales suggests only a one-photon excitation process occurs and carrier-carrier interaction is minimal.

Extrapolation of this trend also suggests our observed dynamics are relevant for amorphization.
The Sb-Te bond distance is around 3.1~\angstrom~\cite{Agafonov:Sb2Te:1991,Zheng:nanores:2016}. 
According to the Lindemann criterion, the solid-to-liquid phase transition occurs when the mean displacement of lattice atoms reaches 10–20\% of the lattice constant~\cite{Sokolowski-Tinten:nature:2003}. 
Thus, we can derive that the solid-to-liquid transition of the crystallized Sb$_{2}$Te sample occurs with a laser excitation fluence of $\approx$ 20~mJ/cm$^{2}$ from the linear relationship observed in \autoref{fig:fluence}(c). 
The predicted melting threshold is close to the observed fluence for laser amorphization of Ge$_{2}$Sb$_{2}$Te$_{5}$ from 14~mJ/cm$^{2}$ and laser ablation from 32~mJ/cm$^{2}$~\cite{Hada:scirep:2015,Waldecker:NM:2015}.

\begin{figure}[!htbp]
	\centering
	\includegraphics[width=0.48\textwidth]{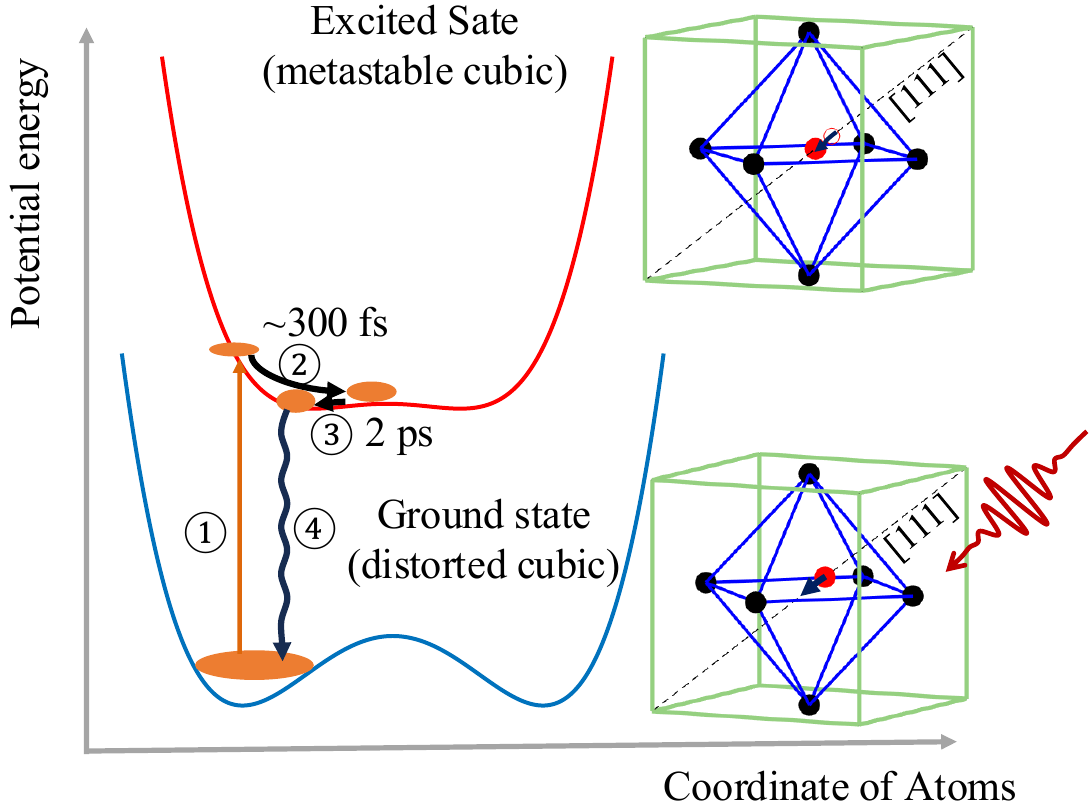}
	\caption{The scheme of the observed ultrafast photo-induced displacive phase transition in a corresponding potential energy change diagram.
    \textcircled{1} shows the electronic excitation, which flattens the double well potential potential energy surface. The distorted Sb atom changes from the minimum position of the ground state potential energy well to the cliff of the flattened excited state potential energy surface. \textcircled{2} shows the distorted Sb atom moves to the bottom of the flattened energy well due to the restoring force. The displacement of the distorted Sb atom to a less distorted (more centered) position happens in a time scale of 300~fs. \textcircled{3} shows the electron-phonon equilibrium process, which happens in a time scale of 2~ps. \textcircled{4} represents the complete nonradiative relaxation of the excited state coordinates back to ground states.
	See text for details. }
	\label{fig:mechansim}
\end{figure}
\autoref{fig:mechansim} shows a physical scheme to explain our observations. The potential energy surface (PES) of Sb$_{2}$Te in the ground state is described by a double well potential due to the Peierls distortion. After the femtosecond pump pulse excitation, charge carriers are generated, which alter the potential energy surface of Sb$_{2}$Te to make it more flat and its curvature shallower~\cite{Fritz:science:2007,Sciaini:nature:2009,Chen:JPCL:2023}. 
Due to the change of PES, it generates a restoring force for the distorted Sb atoms in Sb$_{2}$Te after exciting from the ground state to the electronic excited state~\cite{Chen:PRL:2018}. 
The Sb atoms start the coherent displacement to lower energy in a time scale of around 300~fs as observed in \autoref{fig:fastdyn}(c) top panel.

The electrons and phonons equilibrate in a time constant of around 2~ps, as observed in \autoref{fig:fastdyn}(c) bottom panel. 
The time scale of Sb atomic coherent displacement along [111] to a more centered position matches quite well with the observed resonant bond breaking~\cite{Waldecker:NM:2015,Miller:PRB:2016}. However, the movement of Sb atoms to a more centered octahedral geometry would make the resonant bond stronger~\cite{Kimber:NM:2023}. 
If we look at the experimental difference diffraction pattern at 0.6~ps, as shown in~\autoref{fig:setup}(c), there is an ultrafast decrease in all diffraction peak intensities, not just the anisotropic intensity.
It indicates an ultrafast decrease of the crystalline long-range order, which is important for aligning the p-orbital to form the so called resonant bond~\cite{Miller:PRB:2016}. 
Therefore, the observed ultrafast decrease in long-range order in the crystalline structure is responsible for the resonant bond breaking and, thus, the optical response.

\section{Conclusion}

Phase change materials based on Ge-Sb-Te alloys utilize the reversible transition between the crystalline and amorphous phases to store and erase information. 
Their ultimate time scale of the phase transformation is limited by the atomic movements following a stimuli. 
Recent results on the PCMs of group 1 suggest that there is a non-thermal structural transition to a metastable cubic structure during the crystalline phase to the amorphous phase transition~\cite{Hu:ACSNano:2015,Matsubara:PRL:2016,Qi:PRL:2022}.
Here, we present the comprehensive ultrafast experimental studies on PCMs of group 2, which feature a faster crystalline speed than the PCMs of group 1. 
The ultrafast electron diffraction and non-linear optical spectroscopy results on crystallized Sb$_{2}$Te suggest that it undergoes coherent atomic movements on the femtosecond time scale after laser excitation. 
This displacive phase transition plays a major role in the initial step of the crystalline-to-amorphous transition, as the lattice collapses when the displacive motion reaches a certain amplitude threshold.
Since the displacive phase transition is non-thermal and occurs on the femtosecond time scale, it significantly reduces energy consumption and increases the transformation speed.

Non-thermal displacive phase transitions have been predicted and observed in the GeTe crystalline phase, which undergoes a conformational structural change after laser excitation~\cite{Chen:JPCL:2023,Hu:ACSNano:2015}. 
Our results demonstrate that the non-thermal ultrafast displacive phase transition does happen in the crystalline Sb$_{2}$Te samples under femtosecond laser excitation. 
This finding was somehow not expected to be extended to non-Ge contained Sb$_{2}$Te sample, since Ge atoms play an important role for the GeTe and Ge$_{2}$Sb$_{2}$Te$_{5}$ non-thermal structural dynamics after femtosecond optical pulse excitation.
Distinguished from the rattling motion~\cite{Matsubara:PRL:2016} that has been proposed for Ge$_{2}$Sb$_{2}$Te$_{5}$ and GeTe under femtosecond laser excitation, the high signal-to-noise ratio in our experiments was well within limits to demonstrate that this process involves a coherent atomic displacement instead of rattling motion.
The diffuse scattering background would increase when the rattling motion starts, as the rattling atoms would contribute greatly to it as observed in \ce{VO2}~\cite{Wall:VO2:2018,De:NP:2023}. 
However, our observed diffuse background rises only thermally due to electron-phonon scattering. 
Thus, the TDS results rule out the rattling motion.
Since the displacive phase transition suppresses the Peierls distortion, which is responsible for the band-gap opening~\cite{Gaspard:Peierls:2016}, there is a high possibility that there is an ultrafast conductivity increasing in the femtosecond time scale due to the collapse of the band gap~\cite{Sun:PNAS:2012}. 
This can be further studied by femtosecond pump-probe THz spectroscopy to track the optical conductivity change~\cite{Neu:THz:2018} or angle-resolved photoemission spectroscopy to directly map the dispersion of the energy band~\cite{Sobota:arpes:2021}.

The observed non-thermal sub-picosecond photo-induced displacive phase transition in Sb$_{2}$Te suggests that it is a good candidate for PCMs operating at THz switching speeds, utilizing the light-induced reversible process and bypassing the melting-quench cycle. Compared to PCMs of group 1, Sb$_{2}$Te has a known faster crystallization rate than PCMs of group 1. Our observed results on the non-thermal sub-picosecond photo-induced displacive phase transition of Sb$_{2}$Te suggest that the laser-induced amorphization of Sb$_{2}$Te can be triggered by the displacive phase transition as observed in GeTe and Ge$_{2}$Sb$_{2}$Te$_{5}$. Thus the laser-induced amorphization of Sb$_{2}$Te can also be as fast as the PCMs of group 1. Both the SET and RESET process using Sb$_{2}$Te can be as fast (or even faster) as PCMs of group 1.

\section{Supplementary Material}
The supplementary material contains details regarding the
 \begin{itemize}
	\itemsep=0pt
	\item Laser and Electron Beam Parameters
	\item Rock salt crystal structure factor calculation
	\item Diffraction peak time trace fitting
	\item Coherent atomic displacement and Debye-Waller time constant fit model
 \end{itemize}

\section{Acknowledgement}
\label{sec:Ack}
We thank K. Sokolowski-Tinten from University of Duisburg-Essen and B. J. Siwick from McGill University for valuable discussions. This work was initially supported by the Max Planck Society and the Natural Sciences and Engineering Research Council of Canada (RJDM). It was further supported by the Deutsche Forschungsgemeinschaft (DFG, German Research Foundation) through Projects A06 and C01 of the Collaborative Research Center SFB 1242 “Non-Equilibrium Dynamics of Condensed Matter in the Time Domain” (Project No. 278162697), through Germany’s Excellence Strategy (EXC 2033 - 390677874 - RESOLV) and through SCHL 384/20 - 1 (Project No. 406129719). Additional support was provided by the European Research Council, i.e., ERC - CoG - 2017 SOLWET (Project No. 772286) to R.K.C. 

\medskip

\noindent\textbf{Data availability:} The data that support the findings of this study are available from the
corresponding authors upon reasonable request.

\bibliography{main}

\begin{thebibliography}{56}%
\makeatletter
\providecommand \@ifxundefined [1]{%
 \@ifx{#1\undefined}
}%
\providecommand \@ifnum [1]{%
 \ifnum #1\expandafter \@firstoftwo
 \else \expandafter \@secondoftwo
 \fi
}%
\providecommand \@ifx [1]{%
 \ifx #1\expandafter \@firstoftwo
 \else \expandafter \@secondoftwo
 \fi
}%
\providecommand \natexlab [1]{#1}%
\providecommand \enquote  [1]{``#1''}%
\providecommand \bibnamefont  [1]{#1}%
\providecommand \bibfnamefont [1]{#1}%
\providecommand \citenamefont [1]{#1}%
\providecommand \href@noop [0]{\@secondoftwo}%
\providecommand \href [0]{\begingroup \@sanitize@url \@href}%
\providecommand \@href[1]{\@@startlink{#1}\@@href}%
\providecommand \@@href[1]{\endgroup#1\@@endlink}%
\providecommand \@sanitize@url [0]{\catcode `\\12\catcode `\$12\catcode `\&12\catcode `\#12\catcode `\^12\catcode `\_12\catcode `\%12\relax}%
\providecommand \@@startlink[1]{}%
\providecommand \@@endlink[0]{}%
\providecommand \url  [0]{\begingroup\@sanitize@url \@url }%
\providecommand \@url [1]{\endgroup\@href {#1}{\urlprefix }}%
\providecommand \urlprefix  [0]{URL }%
\providecommand \Eprint [0]{\href }%
\providecommand \doibase [0]{https://doi.org/}%
\providecommand \selectlanguage [0]{\@gobble}%
\providecommand \bibinfo  [0]{\@secondoftwo}%
\providecommand \bibfield  [0]{\@secondoftwo}%
\providecommand \translation [1]{[#1]}%
\providecommand \BibitemOpen [0]{}%
\providecommand \bibitemStop [0]{}%
\providecommand \bibitemNoStop [0]{.\EOS\space}%
\providecommand \EOS [0]{\spacefactor3000\relax}%
\providecommand \BibitemShut  [1]{\csname bibitem#1\endcsname}%
\let\auto@bib@innerbib\@empty
\bibitem [{\citenamefont {Hegedus}\ and\ \citenamefont {Elliott}(2008)}]{Hegedus:intr:2008}%
  \BibitemOpen
  \bibfield  {author} {\bibinfo {author} {\bibfnamefont {J.}~\bibnamefont {Hegedus}}\ and\ \bibinfo {author} {\bibfnamefont {S.~R.}\ \bibnamefont {Elliott}},\ }\bibfield  {title} {\bibinfo {title} {Microscopic origin of the fast crystallization ability of \ce{Ge-Sb-Te} phase-change memory materials},\ }\href {https://doi.org/10.1038/nmat2157} {\bibfield  {journal} {\bibinfo  {journal} {Nat Mater}\ }\textbf {\bibinfo {volume} {7}},\ \bibinfo {pages} {399} (\bibinfo {year} {2008})}\BibitemShut {NoStop}%
\bibitem [{\citenamefont {Zhang}\ \emph {et~al.}(2019)\citenamefont {Zhang}, \citenamefont {Mazzarello}, \citenamefont {Wuttig},\ and\ \citenamefont {Ma}}]{Zhang:Intr:2019}%
  \BibitemOpen
  \bibfield  {author} {\bibinfo {author} {\bibfnamefont {W.}~\bibnamefont {Zhang}}, \bibinfo {author} {\bibfnamefont {R.}~\bibnamefont {Mazzarello}}, \bibinfo {author} {\bibfnamefont {M.}~\bibnamefont {Wuttig}},\ and\ \bibinfo {author} {\bibfnamefont {E.}~\bibnamefont {Ma}},\ }\bibfield  {title} {\bibinfo {title} {Designing crystallization in phase-change materials for universal memory and neuro-inspired computing},\ }\href {https://doi.org/10.1038/s41578-018-0076-x} {\bibfield  {journal} {\bibinfo  {journal} {Nature Reviews Materials}\ }\textbf {\bibinfo {volume} {4}},\ \bibinfo {pages} {150} (\bibinfo {year} {2019})}\BibitemShut {NoStop}%
\bibitem [{\citenamefont {Wuttig}(2005)}]{Wuttg:Intr:2005}%
  \BibitemOpen
  \bibfield  {author} {\bibinfo {author} {\bibfnamefont {M.}~\bibnamefont {Wuttig}},\ }\bibfield  {title} {\bibinfo {title} {Phase-change materials: towards a universal memory?},\ }\href {https://doi.org/10.1038/nmat1359} {\bibfield  {journal} {\bibinfo  {journal} {Nat Mater}\ }\textbf {\bibinfo {volume} {4}},\ \bibinfo {pages} {265} (\bibinfo {year} {2005})}\BibitemShut {NoStop}%
\bibitem [{\citenamefont {Lankhorst}\ \emph {et~al.}(2005)\citenamefont {Lankhorst}, \citenamefont {Ketelaars},\ and\ \citenamefont {Wolters}}]{Lankhorst:Intr:2005}%
  \BibitemOpen
  \bibfield  {author} {\bibinfo {author} {\bibfnamefont {M.~H.~R.}\ \bibnamefont {Lankhorst}}, \bibinfo {author} {\bibfnamefont {B.~W. S. M.~M.}\ \bibnamefont {Ketelaars}},\ and\ \bibinfo {author} {\bibfnamefont {R.~A.~M.}\ \bibnamefont {Wolters}},\ }\bibfield  {title} {\bibinfo {title} {Low-cost and nanoscale non-volatile memory concept for future silicon chips},\ }\href {https://doi.org/10.1038/nmat1350} {\bibfield  {journal} {\bibinfo  {journal} {Nature Materials}\ }\textbf {\bibinfo {volume} {4}},\ \bibinfo {pages} {347} (\bibinfo {year} {2005})}\BibitemShut {NoStop}%
\bibitem [{\citenamefont {Atwood}(2008)}]{Atwood:Intr:2008}%
  \BibitemOpen
  \bibfield  {author} {\bibinfo {author} {\bibfnamefont {G.}~\bibnamefont {Atwood}},\ }\bibfield  {title} {\bibinfo {title} {Engineering. phase-change materials for electronic memories},\ }\href {https://doi.org/10.1126/science.1160231} {\bibfield  {journal} {\bibinfo  {journal} {Science}\ }\textbf {\bibinfo {volume} {321}},\ \bibinfo {pages} {210} (\bibinfo {year} {2008})}\BibitemShut {NoStop}%
\bibitem [{\citenamefont {Lotnyk}\ \emph {et~al.}(2019)\citenamefont {Lotnyk}, \citenamefont {Behrens},\ and\ \citenamefont {Rauschenbach}}]{Lotnyk:Intr:2019}%
  \BibitemOpen
  \bibfield  {author} {\bibinfo {author} {\bibfnamefont {A.}~\bibnamefont {Lotnyk}}, \bibinfo {author} {\bibfnamefont {M.}~\bibnamefont {Behrens}},\ and\ \bibinfo {author} {\bibfnamefont {B.}~\bibnamefont {Rauschenbach}},\ }\bibfield  {title} {\bibinfo {title} {Phase change thin films for non-volatile memory applications},\ }\href {https://doi.org/10.1039/c9na00366e} {\bibfield  {journal} {\bibinfo  {journal} {Nanoscale Adv}\ }\textbf {\bibinfo {volume} {1}},\ \bibinfo {pages} {3836} (\bibinfo {year} {2019})}\BibitemShut {NoStop}%
\bibitem [{\citenamefont {Kolobov}\ \emph {et~al.}(2011)\citenamefont {Kolobov}, \citenamefont {Krbal}, \citenamefont {Fons}, \citenamefont {Tominaga},\ and\ \citenamefont {Uruga}}]{Kolobov:NC:2011}%
  \BibitemOpen
  \bibfield  {author} {\bibinfo {author} {\bibfnamefont {A.~V.}\ \bibnamefont {Kolobov}}, \bibinfo {author} {\bibfnamefont {M.}~\bibnamefont {Krbal}}, \bibinfo {author} {\bibfnamefont {P.}~\bibnamefont {Fons}}, \bibinfo {author} {\bibfnamefont {J.}~\bibnamefont {Tominaga}},\ and\ \bibinfo {author} {\bibfnamefont {T.}~\bibnamefont {Uruga}},\ }\bibfield  {title} {\bibinfo {title} {Distortion-triggered loss of long-range order in solids with bonding energy hierarchy},\ }\href {https://doi.org/10.1038/nchem.1007} {\bibfield  {journal} {\bibinfo  {journal} {Nat Chem}\ }\textbf {\bibinfo {volume} {3}},\ \bibinfo {pages} {311} (\bibinfo {year} {2011})}\BibitemShut {NoStop}%
\bibitem [{\citenamefont {Nam}\ \emph {et~al.}(2012)\citenamefont {Nam}, \citenamefont {Chung}, \citenamefont {Lo}, \citenamefont {Qi}, \citenamefont {Li}, \citenamefont {Lu}, \citenamefont {Johnson}, \citenamefont {Jung}, \citenamefont {Nukala},\ and\ \citenamefont {Agarwal}}]{Nam:science:2012}%
  \BibitemOpen
  \bibfield  {author} {\bibinfo {author} {\bibfnamefont {S.~W.}\ \bibnamefont {Nam}}, \bibinfo {author} {\bibfnamefont {H.~S.}\ \bibnamefont {Chung}}, \bibinfo {author} {\bibfnamefont {Y.~C.}\ \bibnamefont {Lo}}, \bibinfo {author} {\bibfnamefont {L.}~\bibnamefont {Qi}}, \bibinfo {author} {\bibfnamefont {J.}~\bibnamefont {Li}}, \bibinfo {author} {\bibfnamefont {Y.}~\bibnamefont {Lu}}, \bibinfo {author} {\bibfnamefont {A.~T.~C.}\ \bibnamefont {Johnson}}, \bibinfo {author} {\bibfnamefont {Y.~W.}\ \bibnamefont {Jung}}, \bibinfo {author} {\bibfnamefont {P.}~\bibnamefont {Nukala}},\ and\ \bibinfo {author} {\bibfnamefont {R.}~\bibnamefont {Agarwal}},\ }\bibfield  {title} {\bibinfo {title} {Electrical wind force-driven and dislocation-templated amorphization in phase-change nanowires},\ }\href {https://doi.org/10.1126/science.1220119} {\bibfield  {journal} {\bibinfo  {journal} {Science}\ }\textbf {\bibinfo {volume} {336}},\ \bibinfo {pages} {1561} (\bibinfo {year} {2012})}\BibitemShut {NoStop}%
\bibitem [{\citenamefont {Loke}\ \emph {et~al.}(2012)\citenamefont {Loke}, \citenamefont {Lee}, \citenamefont {Wang}, \citenamefont {Shi}, \citenamefont {Zhao}, \citenamefont {Yeo}, \citenamefont {Chong},\ and\ \citenamefont {Elliott}}]{Loke:Intr:2012}%
  \BibitemOpen
  \bibfield  {author} {\bibinfo {author} {\bibfnamefont {D.}~\bibnamefont {Loke}}, \bibinfo {author} {\bibfnamefont {T.~H.}\ \bibnamefont {Lee}}, \bibinfo {author} {\bibfnamefont {W.~J.}\ \bibnamefont {Wang}}, \bibinfo {author} {\bibfnamefont {L.~P.}\ \bibnamefont {Shi}}, \bibinfo {author} {\bibfnamefont {R.}~\bibnamefont {Zhao}}, \bibinfo {author} {\bibfnamefont {Y.~C.}\ \bibnamefont {Yeo}}, \bibinfo {author} {\bibfnamefont {T.~C.}\ \bibnamefont {Chong}},\ and\ \bibinfo {author} {\bibfnamefont {S.~R.}\ \bibnamefont {Elliott}},\ }\bibfield  {title} {\bibinfo {title} {Breaking the speed limits of phase-change memory},\ }\href {https://doi.org/10.1126/science.1221561} {\bibfield  {journal} {\bibinfo  {journal} {Science}\ }\textbf {\bibinfo {volume} {336}},\ \bibinfo {pages} {1566} (\bibinfo {year} {2012})}\BibitemShut {NoStop}%
\bibitem [{\citenamefont {Rao}\ \emph {et~al.}(2017)\citenamefont {Rao}, \citenamefont {Ding}, \citenamefont {Zhou}, \citenamefont {Zheng}, \citenamefont {Xia}, \citenamefont {Lv}, \citenamefont {Song}, \citenamefont {Feng}, \citenamefont {Ronneberger}, \citenamefont {Mazzarello}, \citenamefont {Zhang},\ and\ \citenamefont {Ma}}]{Rao:science:2017}%
  \BibitemOpen
  \bibfield  {author} {\bibinfo {author} {\bibfnamefont {F.}~\bibnamefont {Rao}}, \bibinfo {author} {\bibfnamefont {K.}~\bibnamefont {Ding}}, \bibinfo {author} {\bibfnamefont {Y.}~\bibnamefont {Zhou}}, \bibinfo {author} {\bibfnamefont {Y.}~\bibnamefont {Zheng}}, \bibinfo {author} {\bibfnamefont {M.}~\bibnamefont {Xia}}, \bibinfo {author} {\bibfnamefont {S.}~\bibnamefont {Lv}}, \bibinfo {author} {\bibfnamefont {Z.}~\bibnamefont {Song}}, \bibinfo {author} {\bibfnamefont {S.}~\bibnamefont {Feng}}, \bibinfo {author} {\bibfnamefont {I.}~\bibnamefont {Ronneberger}}, \bibinfo {author} {\bibfnamefont {R.}~\bibnamefont {Mazzarello}}, \bibinfo {author} {\bibfnamefont {W.}~\bibnamefont {Zhang}},\ and\ \bibinfo {author} {\bibfnamefont {E.}~\bibnamefont {Ma}},\ }\bibfield  {title} {\bibinfo {title} {Reducing the stochasticity of crystal nucleation to enable subnanosecond memory writing},\ }\href {https://doi.org/10.1126/science.aao3212} {\bibfield  {journal} {\bibinfo  {journal} {Science}\ }\textbf {\bibinfo {volume} {358}},\ \bibinfo {pages} {1423} (\bibinfo {year} {2017})}\BibitemShut {NoStop}%
\bibitem [{\citenamefont {Edwards}\ \emph {et~al.}(2006)\citenamefont {Edwards}, \citenamefont {Pineda}, \citenamefont {Schultz}, \citenamefont {Martin}, \citenamefont {Thompson}, \citenamefont {Hjalmarson},\ and\ \citenamefont {Umrigar}}]{Edwards:PRB:2006}%
  \BibitemOpen
  \bibfield  {author} {\bibinfo {author} {\bibfnamefont {A.}~\bibnamefont {Edwards}}, \bibinfo {author} {\bibfnamefont {A.}~\bibnamefont {Pineda}}, \bibinfo {author} {\bibfnamefont {P.}~\bibnamefont {Schultz}}, \bibinfo {author} {\bibfnamefont {M.}~\bibnamefont {Martin}}, \bibinfo {author} {\bibfnamefont {A.}~\bibnamefont {Thompson}}, \bibinfo {author} {\bibfnamefont {H.}~\bibnamefont {Hjalmarson}},\ and\ \bibinfo {author} {\bibfnamefont {C.}~\bibnamefont {Umrigar}},\ }\bibfield  {title} {\bibinfo {title} {Electronic structure of intrinsic defects in crystalline germanium telluride},\ }\bibfield  {journal} {\bibinfo  {journal} {Physical Review B}\ }\textbf {\bibinfo {volume} {73}},\ \href {https://doi.org/10.1103/PhysRevB.73.045210} {10.1103/PhysRevB.73.045210} (\bibinfo {year} {2006})\BibitemShut {NoStop}%
\bibitem [{\citenamefont {Fons}\ \emph {et~al.}(2010)\citenamefont {Fons}, \citenamefont {Osawa}, \citenamefont {Kolobov}, \citenamefont {Fukaya}, \citenamefont {Suzuki}, \citenamefont {Uruga}, \citenamefont {Kawamura}, \citenamefont {Tanida},\ and\ \citenamefont {Tominaga}}]{Fons:PRB:2010}%
  \BibitemOpen
  \bibfield  {author} {\bibinfo {author} {\bibfnamefont {P.}~\bibnamefont {Fons}}, \bibinfo {author} {\bibfnamefont {H.}~\bibnamefont {Osawa}}, \bibinfo {author} {\bibfnamefont {A.~V.}\ \bibnamefont {Kolobov}}, \bibinfo {author} {\bibfnamefont {T.}~\bibnamefont {Fukaya}}, \bibinfo {author} {\bibfnamefont {M.}~\bibnamefont {Suzuki}}, \bibinfo {author} {\bibfnamefont {T.}~\bibnamefont {Uruga}}, \bibinfo {author} {\bibfnamefont {N.}~\bibnamefont {Kawamura}}, \bibinfo {author} {\bibfnamefont {H.}~\bibnamefont {Tanida}},\ and\ \bibinfo {author} {\bibfnamefont {J.}~\bibnamefont {Tominaga}},\ }\bibfield  {title} {\bibinfo {title} {Photoassisted amorphization of the phase-change memory alloy \ce{Ge2Sb2Te5}},\ }\bibfield  {journal} {\bibinfo  {journal} {Physical Review B}\ }\textbf {\bibinfo {volume} {82}},\ \href {https://doi.org/10.1103/PhysRevB.82.041203} {10.1103/PhysRevB.82.041203} (\bibinfo {year} {2010})\BibitemShut {NoStop}%
\bibitem [{\citenamefont {Mukhopadhyay}\ \emph {et~al.}(2016)\citenamefont {Mukhopadhyay}, \citenamefont {Sun}, \citenamefont {Subedi}, \citenamefont {Siegrist},\ and\ \citenamefont {Singh}}]{Mukhopadhyay:scirep:2016}%
  \BibitemOpen
  \bibfield  {author} {\bibinfo {author} {\bibfnamefont {S.}~\bibnamefont {Mukhopadhyay}}, \bibinfo {author} {\bibfnamefont {J.}~\bibnamefont {Sun}}, \bibinfo {author} {\bibfnamefont {A.}~\bibnamefont {Subedi}}, \bibinfo {author} {\bibfnamefont {T.}~\bibnamefont {Siegrist}},\ and\ \bibinfo {author} {\bibfnamefont {D.~J.}\ \bibnamefont {Singh}},\ }\bibfield  {title} {\bibinfo {title} {Competing covalent and ionic bonding in \ce{Ge-Sb-Te} phase change materials},\ }\href {https://doi.org/10.1038/srep25981} {\bibfield  {journal} {\bibinfo  {journal} {Sci Rep}\ }\textbf {\bibinfo {volume} {6}},\ \bibinfo {pages} {25981} (\bibinfo {year} {2016})}\BibitemShut {NoStop}%
\bibitem [{\citenamefont {Mott}(1969)}]{Mott:1969}%
  \BibitemOpen
  \bibfield  {author} {\bibinfo {author} {\bibfnamefont {N.~F.}\ \bibnamefont {Mott}},\ }\bibfield  {title} {\bibinfo {title} {Conduction in non-crystalline materials},\ }\href {https://doi.org/10.1080/14786436908216338} {\bibfield  {journal} {\bibinfo  {journal} {Philosophical Magazine}\ }\textbf {\bibinfo {volume} {19}},\ \bibinfo {pages} {835} (\bibinfo {year} {1969})}\BibitemShut {NoStop}%
\bibitem [{\citenamefont {Robertson}(2016)}]{Robertson:bonding:2016}%
  \BibitemOpen
  \bibfield  {author} {\bibinfo {author} {\bibfnamefont {J.}~\bibnamefont {Robertson}},\ }\bibfield  {title} {\bibinfo {title} {Mott lecture: How bonding concepts can help understand amorphous semiconductor behavior},\ }\href {https://doi.org/10.1002/pssa.201532875} {\bibfield  {journal} {\bibinfo  {journal} {physica status solidi (a)}\ }\textbf {\bibinfo {volume} {213}},\ \bibinfo {pages} {1641} (\bibinfo {year} {2016})}\BibitemShut {NoStop}%
\bibitem [{\citenamefont {Shportko}\ \emph {et~al.}(2008)\citenamefont {Shportko}, \citenamefont {Kremers}, \citenamefont {Woda}, \citenamefont {Lencer}, \citenamefont {Robertson},\ and\ \citenamefont {Wuttig}}]{Shportko:NM:2008}%
  \BibitemOpen
  \bibfield  {author} {\bibinfo {author} {\bibfnamefont {K.}~\bibnamefont {Shportko}}, \bibinfo {author} {\bibfnamefont {S.}~\bibnamefont {Kremers}}, \bibinfo {author} {\bibfnamefont {M.}~\bibnamefont {Woda}}, \bibinfo {author} {\bibfnamefont {D.}~\bibnamefont {Lencer}}, \bibinfo {author} {\bibfnamefont {J.}~\bibnamefont {Robertson}},\ and\ \bibinfo {author} {\bibfnamefont {M.}~\bibnamefont {Wuttig}},\ }\bibfield  {title} {\bibinfo {title} {Resonant bonding in crystalline phase-change materials},\ }\href {https://doi.org/10.1038/nmat2226} {\bibfield  {journal} {\bibinfo  {journal} {Nat Mater}\ }\textbf {\bibinfo {volume} {7}},\ \bibinfo {pages} {653} (\bibinfo {year} {2008})}\BibitemShut {NoStop}%
\bibitem [{\citenamefont {Matsunaga}\ \emph {et~al.}(2011)\citenamefont {Matsunaga}, \citenamefont {Akola}, \citenamefont {Kohara}, \citenamefont {Honma}, \citenamefont {Kobayashi}, \citenamefont {Ikenaga}, \citenamefont {Jones}, \citenamefont {Yamada}, \citenamefont {Takata},\ and\ \citenamefont {Kojima}}]{Matsunaga:NM:2011}%
  \BibitemOpen
  \bibfield  {author} {\bibinfo {author} {\bibfnamefont {T.}~\bibnamefont {Matsunaga}}, \bibinfo {author} {\bibfnamefont {J.}~\bibnamefont {Akola}}, \bibinfo {author} {\bibfnamefont {S.}~\bibnamefont {Kohara}}, \bibinfo {author} {\bibfnamefont {T.}~\bibnamefont {Honma}}, \bibinfo {author} {\bibfnamefont {K.}~\bibnamefont {Kobayashi}}, \bibinfo {author} {\bibfnamefont {E.}~\bibnamefont {Ikenaga}}, \bibinfo {author} {\bibfnamefont {R.~O.}\ \bibnamefont {Jones}}, \bibinfo {author} {\bibfnamefont {N.}~\bibnamefont {Yamada}}, \bibinfo {author} {\bibfnamefont {M.}~\bibnamefont {Takata}},\ and\ \bibinfo {author} {\bibfnamefont {R.}~\bibnamefont {Kojima}},\ }\bibfield  {title} {\bibinfo {title} {From local structure to nanosecond recrystallization dynamics in aginsbte phase-change materials},\ }\href {https://doi.org/10.1038/nmat2931} {\bibfield  {journal} {\bibinfo  {journal} {Nat Mater}\ }\textbf {\bibinfo {volume} {10}},\ \bibinfo {pages} {129} (\bibinfo {year} {2011})}\BibitemShut {NoStop}%
\bibitem [{\citenamefont {Shen}\ \emph {et~al.}(2023)\citenamefont {Shen}, \citenamefont {Song}, \citenamefont {Ren}, \citenamefont {Song}, \citenamefont {Zhou},\ and\ \citenamefont {Zhu}}]{Shen:Intr:2023}%
  \BibitemOpen
  \bibfield  {author} {\bibinfo {author} {\bibfnamefont {J.}~\bibnamefont {Shen}}, \bibinfo {author} {\bibfnamefont {W.}~\bibnamefont {Song}}, \bibinfo {author} {\bibfnamefont {K.}~\bibnamefont {Ren}}, \bibinfo {author} {\bibfnamefont {Z.}~\bibnamefont {Song}}, \bibinfo {author} {\bibfnamefont {P.}~\bibnamefont {Zhou}},\ and\ \bibinfo {author} {\bibfnamefont {M.}~\bibnamefont {Zhu}},\ }\bibfield  {title} {\bibinfo {title} {Toward the speed limit of phase-change memory},\ }\href {https://doi.org/10.1002/adma.202208065} {\bibfield  {journal} {\bibinfo  {journal} {Adv Mater}\ }\textbf {\bibinfo {volume} {35}},\ \bibinfo {pages} {e2208065} (\bibinfo {year} {2023})}\BibitemShut {NoStop}%
\bibitem [{\citenamefont {Hu}\ \emph {et~al.}(2015)\citenamefont {Hu}, \citenamefont {Vanacore}, \citenamefont {Yang}, \citenamefont {Miao},\ and\ \citenamefont {Zewail}}]{Hu:ACSNano:2015}%
  \BibitemOpen
  \bibfield  {author} {\bibinfo {author} {\bibfnamefont {J.}~\bibnamefont {Hu}}, \bibinfo {author} {\bibfnamefont {G.~M.}\ \bibnamefont {Vanacore}}, \bibinfo {author} {\bibfnamefont {Z.}~\bibnamefont {Yang}}, \bibinfo {author} {\bibfnamefont {X.}~\bibnamefont {Miao}},\ and\ \bibinfo {author} {\bibfnamefont {A.~H.}\ \bibnamefont {Zewail}},\ }\bibfield  {title} {\bibinfo {title} {Transient structures and possible limits of data recording in phase-change materials},\ }\href {https://doi.org/10.1021/acsnano.5b01965} {\bibfield  {journal} {\bibinfo  {journal} {ACS Nano}\ }\textbf {\bibinfo {volume} {9}},\ \bibinfo {pages} {6728} (\bibinfo {year} {2015})}\BibitemShut {NoStop}%
\bibitem [{\citenamefont {Waldecker}\ \emph {et~al.}(2015)\citenamefont {Waldecker}, \citenamefont {Miller}, \citenamefont {Rude}, \citenamefont {Bertoni}, \citenamefont {Osmond}, \citenamefont {Pruneri}, \citenamefont {Simpson}, \citenamefont {Ernstorfer},\ and\ \citenamefont {Wall}}]{Waldecker:NM:2015}%
  \BibitemOpen
  \bibfield  {author} {\bibinfo {author} {\bibfnamefont {L.}~\bibnamefont {Waldecker}}, \bibinfo {author} {\bibfnamefont {T.~A.}\ \bibnamefont {Miller}}, \bibinfo {author} {\bibfnamefont {M.}~\bibnamefont {Rude}}, \bibinfo {author} {\bibfnamefont {R.}~\bibnamefont {Bertoni}}, \bibinfo {author} {\bibfnamefont {J.}~\bibnamefont {Osmond}}, \bibinfo {author} {\bibfnamefont {V.}~\bibnamefont {Pruneri}}, \bibinfo {author} {\bibfnamefont {R.~E.}\ \bibnamefont {Simpson}}, \bibinfo {author} {\bibfnamefont {R.}~\bibnamefont {Ernstorfer}},\ and\ \bibinfo {author} {\bibfnamefont {S.}~\bibnamefont {Wall}},\ }\bibfield  {title} {\bibinfo {title} {Time-domain separation of optical properties from structural transitions in resonantly bonded materials},\ }\href {https://doi.org/10.1038/nmat4359} {\bibfield  {journal} {\bibinfo  {journal} {Nat Mater}\ }\textbf {\bibinfo {volume} {14}},\ \bibinfo {pages} {991} (\bibinfo {year} {2015})}\BibitemShut {NoStop}%
\bibitem [{\citenamefont {Hada}\ \emph {et~al.}(2015)\citenamefont {Hada}, \citenamefont {Oba}, \citenamefont {Kuwahara}, \citenamefont {Katayama}, \citenamefont {Saiki}, \citenamefont {Takeda},\ and\ \citenamefont {Nakamura}}]{Hada:scirep:2015}%
  \BibitemOpen
  \bibfield  {author} {\bibinfo {author} {\bibfnamefont {M.}~\bibnamefont {Hada}}, \bibinfo {author} {\bibfnamefont {W.}~\bibnamefont {Oba}}, \bibinfo {author} {\bibfnamefont {M.}~\bibnamefont {Kuwahara}}, \bibinfo {author} {\bibfnamefont {I.}~\bibnamefont {Katayama}}, \bibinfo {author} {\bibfnamefont {T.}~\bibnamefont {Saiki}}, \bibinfo {author} {\bibfnamefont {J.}~\bibnamefont {Takeda}},\ and\ \bibinfo {author} {\bibfnamefont {K.~G.}\ \bibnamefont {Nakamura}},\ }\bibfield  {title} {\bibinfo {title} {Ultrafast time-resolved electron diffraction revealing the nonthermal dynamics of near-uv photoexcitation-induced amorphization in \ce{Ge2Sb2Te5}},\ }\href {https://doi.org/10.1038/srep13530} {\bibfield  {journal} {\bibinfo  {journal} {Sci Rep}\ }\textbf {\bibinfo {volume} {5}},\ \bibinfo {pages} {13530} (\bibinfo {year} {2015})}\BibitemShut {NoStop}%
\bibitem [{\citenamefont {Matsubara}\ \emph {et~al.}(2016)\citenamefont {Matsubara}, \citenamefont {Okada}, \citenamefont {Ichitsubo}, \citenamefont {Kawaguchi}, \citenamefont {Hirata}, \citenamefont {Guan}, \citenamefont {Tokuda}, \citenamefont {Tanimura}, \citenamefont {Matsunaga}, \citenamefont {Chen},\ and\ \citenamefont {Yamada}}]{Matsubara:PRL:2016}%
  \BibitemOpen
  \bibfield  {author} {\bibinfo {author} {\bibfnamefont {E.}~\bibnamefont {Matsubara}}, \bibinfo {author} {\bibfnamefont {S.}~\bibnamefont {Okada}}, \bibinfo {author} {\bibfnamefont {T.}~\bibnamefont {Ichitsubo}}, \bibinfo {author} {\bibfnamefont {T.}~\bibnamefont {Kawaguchi}}, \bibinfo {author} {\bibfnamefont {A.}~\bibnamefont {Hirata}}, \bibinfo {author} {\bibfnamefont {P.~F.}\ \bibnamefont {Guan}}, \bibinfo {author} {\bibfnamefont {K.}~\bibnamefont {Tokuda}}, \bibinfo {author} {\bibfnamefont {K.}~\bibnamefont {Tanimura}}, \bibinfo {author} {\bibfnamefont {T.}~\bibnamefont {Matsunaga}}, \bibinfo {author} {\bibfnamefont {M.~W.}\ \bibnamefont {Chen}},\ and\ \bibinfo {author} {\bibfnamefont {N.}~\bibnamefont {Yamada}},\ }\bibfield  {title} {\bibinfo {title} {Initial atomic motion immediately following femtosecond-laser excitation in phase-change materials},\ }\href {https://doi.org/10.1103/PhysRevLett.117.135501} {\bibfield  {journal} {\bibinfo  {journal} {Phys Rev Lett}\ }\textbf {\bibinfo {volume} {117}},\ \bibinfo {pages} {135501} (\bibinfo {year} {2016})}\BibitemShut {NoStop}%
\bibitem [{\citenamefont {Mitrofanov}\ \emph {et~al.}(2016)\citenamefont {Mitrofanov}, \citenamefont {Fons}, \citenamefont {Makino}, \citenamefont {Terashima}, \citenamefont {Shimada}, \citenamefont {Kolobov}, \citenamefont {Tominaga}, \citenamefont {Bragaglia}, \citenamefont {Giussani}, \citenamefont {Calarco}, \citenamefont {Riechert}, \citenamefont {Sato}, \citenamefont {Katayama}, \citenamefont {Ogawa}, \citenamefont {Togashi}, \citenamefont {Yabashi}, \citenamefont {Wall}, \citenamefont {Brewe},\ and\ \citenamefont {Hase}}]{Mitrofanov:scirep:2016}%
  \BibitemOpen
  \bibfield  {author} {\bibinfo {author} {\bibfnamefont {K.~V.}\ \bibnamefont {Mitrofanov}}, \bibinfo {author} {\bibfnamefont {P.}~\bibnamefont {Fons}}, \bibinfo {author} {\bibfnamefont {K.}~\bibnamefont {Makino}}, \bibinfo {author} {\bibfnamefont {R.}~\bibnamefont {Terashima}}, \bibinfo {author} {\bibfnamefont {T.}~\bibnamefont {Shimada}}, \bibinfo {author} {\bibfnamefont {A.~V.}\ \bibnamefont {Kolobov}}, \bibinfo {author} {\bibfnamefont {J.}~\bibnamefont {Tominaga}}, \bibinfo {author} {\bibfnamefont {V.}~\bibnamefont {Bragaglia}}, \bibinfo {author} {\bibfnamefont {A.}~\bibnamefont {Giussani}}, \bibinfo {author} {\bibfnamefont {R.}~\bibnamefont {Calarco}}, \bibinfo {author} {\bibfnamefont {H.}~\bibnamefont {Riechert}}, \bibinfo {author} {\bibfnamefont {T.}~\bibnamefont {Sato}}, \bibinfo {author} {\bibfnamefont {T.}~\bibnamefont {Katayama}}, \bibinfo {author} {\bibfnamefont {K.}~\bibnamefont {Ogawa}}, \bibinfo {author} {\bibfnamefont {T.}~\bibnamefont {Togashi}}, \bibinfo {author} {\bibfnamefont {M.}~\bibnamefont {Yabashi}}, \bibinfo {author} {\bibfnamefont {S.}~\bibnamefont {Wall}}, \bibinfo {author} {\bibfnamefont {D.}~\bibnamefont {Brewe}},\ and\ \bibinfo {author} {\bibfnamefont {M.}~\bibnamefont {Hase}},\ }\bibfield  {title} {\bibinfo {title} {Sub-nanometre resolution of atomic motion during electronic excitation in phase-change materials},\ }\href {https://doi.org/10.1038/srep20633} {\bibfield  {journal} {\bibinfo  {journal} {Sci Rep}\ }\textbf {\bibinfo {volume} {6}},\ \bibinfo {pages} {20633} (\bibinfo {year} {2016})}\BibitemShut {NoStop}%
\bibitem [{\citenamefont {Zalden}\ \emph {et~al.}(2019)\citenamefont {Zalden}, \citenamefont {Quirin}, \citenamefont {Schumacher}, \citenamefont {Siegel}, \citenamefont {Wei}, \citenamefont {Koc}, \citenamefont {Nicoul}, \citenamefont {Trigo}, \citenamefont {Andreasson}, \citenamefont {Enquist}, \citenamefont {Shu}, \citenamefont {Pardini}, \citenamefont {Chollet}, \citenamefont {Zhu}, \citenamefont {Lemke}, \citenamefont {Ronneberger}, \citenamefont {Larsson}, \citenamefont {Lindenberg}, \citenamefont {Fischer}, \citenamefont {Hau-Riege}, \citenamefont {Reis}, \citenamefont {Mazzarello}, \citenamefont {Wuttig},\ and\ \citenamefont {Sokolowski-Tinten}}]{Zalden:science:2019}%
  \BibitemOpen
  \bibfield  {author} {\bibinfo {author} {\bibfnamefont {P.}~\bibnamefont {Zalden}}, \bibinfo {author} {\bibfnamefont {F.}~\bibnamefont {Quirin}}, \bibinfo {author} {\bibfnamefont {M.}~\bibnamefont {Schumacher}}, \bibinfo {author} {\bibfnamefont {J.}~\bibnamefont {Siegel}}, \bibinfo {author} {\bibfnamefont {S.}~\bibnamefont {Wei}}, \bibinfo {author} {\bibfnamefont {A.}~\bibnamefont {Koc}}, \bibinfo {author} {\bibfnamefont {M.}~\bibnamefont {Nicoul}}, \bibinfo {author} {\bibfnamefont {M.}~\bibnamefont {Trigo}}, \bibinfo {author} {\bibfnamefont {P.}~\bibnamefont {Andreasson}}, \bibinfo {author} {\bibfnamefont {H.}~\bibnamefont {Enquist}}, \bibinfo {author} {\bibfnamefont {M.~J.}\ \bibnamefont {Shu}}, \bibinfo {author} {\bibfnamefont {T.}~\bibnamefont {Pardini}}, \bibinfo {author} {\bibfnamefont {M.}~\bibnamefont {Chollet}}, \bibinfo {author} {\bibfnamefont {D.}~\bibnamefont {Zhu}}, \bibinfo {author} {\bibfnamefont {H.}~\bibnamefont {Lemke}}, \bibinfo {author} {\bibfnamefont {I.}~\bibnamefont {Ronneberger}}, \bibinfo {author} {\bibfnamefont {J.}~\bibnamefont {Larsson}}, \bibinfo {author} {\bibfnamefont {A.~M.}\ \bibnamefont {Lindenberg}}, \bibinfo {author} {\bibfnamefont {H.~E.}\ \bibnamefont {Fischer}}, \bibinfo {author} {\bibfnamefont {S.}~\bibnamefont {Hau-Riege}}, \bibinfo {author} {\bibfnamefont {D.~A.}\ \bibnamefont {Reis}}, \bibinfo {author} {\bibfnamefont {R.}~\bibnamefont {Mazzarello}}, \bibinfo {author} {\bibfnamefont {M.}~\bibnamefont {Wuttig}},\ and\ \bibinfo {author} {\bibfnamefont {K.}~\bibnamefont {Sokolowski-Tinten}},\ }\bibfield  {title} {\bibinfo {title} {Femtosecond x-ray diffraction reveals a liquid-liquid phase transition in phase-change materials},\ }\href {https://doi.org/10.1126/science.aaw1773} {\bibfield  {journal} {\bibinfo  {journal} {Science}\ }\textbf {\bibinfo {volume} {364}},\ \bibinfo {pages} {1062} (\bibinfo {year} {2019})}\BibitemShut {NoStop}%
\bibitem [{\citenamefont {Qi}\ \emph {et~al.}(2022)\citenamefont {Qi}, \citenamefont {Chen}, \citenamefont {Vasileiadis}, \citenamefont {Zahn}, \citenamefont {Seiler}, \citenamefont {Li},\ and\ \citenamefont {Ernstorfer}}]{Qi:PRL:2022}%
  \BibitemOpen
  \bibfield  {author} {\bibinfo {author} {\bibfnamefont {Y.}~\bibnamefont {Qi}}, \bibinfo {author} {\bibfnamefont {N.}~\bibnamefont {Chen}}, \bibinfo {author} {\bibfnamefont {T.}~\bibnamefont {Vasileiadis}}, \bibinfo {author} {\bibfnamefont {D.}~\bibnamefont {Zahn}}, \bibinfo {author} {\bibfnamefont {H.}~\bibnamefont {Seiler}}, \bibinfo {author} {\bibfnamefont {X.}~\bibnamefont {Li}},\ and\ \bibinfo {author} {\bibfnamefont {R.}~\bibnamefont {Ernstorfer}},\ }\bibfield  {title} {\bibinfo {title} {Photoinduced ultrafast transition of the local correlated structure in chalcogenide phase-change materials},\ }\href {https://doi.org/10.1103/PhysRevLett.129.135701} {\bibfield  {journal} {\bibinfo  {journal} {Phys Rev Lett}\ }\textbf {\bibinfo {volume} {129}},\ \bibinfo {pages} {135701} (\bibinfo {year} {2022})}\BibitemShut {NoStop}%
\bibitem [{\citenamefont {Kolobov}\ \emph {et~al.}(2004)\citenamefont {Kolobov}, \citenamefont {Fons}, \citenamefont {Frenkel}, \citenamefont {Ankudinov}, \citenamefont {Tominaga},\ and\ \citenamefont {Uruga}}]{Kolobov:NM:2004}%
  \BibitemOpen
  \bibfield  {author} {\bibinfo {author} {\bibfnamefont {A.~V.}\ \bibnamefont {Kolobov}}, \bibinfo {author} {\bibfnamefont {P.}~\bibnamefont {Fons}}, \bibinfo {author} {\bibfnamefont {A.~I.}\ \bibnamefont {Frenkel}}, \bibinfo {author} {\bibfnamefont {A.~L.}\ \bibnamefont {Ankudinov}}, \bibinfo {author} {\bibfnamefont {J.}~\bibnamefont {Tominaga}},\ and\ \bibinfo {author} {\bibfnamefont {T.}~\bibnamefont {Uruga}},\ }\bibfield  {title} {\bibinfo {title} {Understanding the phase-change mechanism of rewritable optical media},\ }\href {https://doi.org/10.1038/nmat1215} {\bibfield  {journal} {\bibinfo  {journal} {Nat Mater}\ }\textbf {\bibinfo {volume} {3}},\ \bibinfo {pages} {703} (\bibinfo {year} {2004})}\BibitemShut {NoStop}%
\bibitem [{\citenamefont {Chen}\ \emph {et~al.}(2018)\citenamefont {Chen}, \citenamefont {Li}, \citenamefont {Bang}, \citenamefont {Wang}, \citenamefont {Han}, \citenamefont {West}, \citenamefont {Zhang},\ and\ \citenamefont {Sun}}]{Chen:PRL:2018}%
  \BibitemOpen
  \bibfield  {author} {\bibinfo {author} {\bibfnamefont {N.~K.}\ \bibnamefont {Chen}}, \bibinfo {author} {\bibfnamefont {X.~B.}\ \bibnamefont {Li}}, \bibinfo {author} {\bibfnamefont {J.}~\bibnamefont {Bang}}, \bibinfo {author} {\bibfnamefont {X.~P.}\ \bibnamefont {Wang}}, \bibinfo {author} {\bibfnamefont {D.}~\bibnamefont {Han}}, \bibinfo {author} {\bibfnamefont {D.}~\bibnamefont {West}}, \bibinfo {author} {\bibfnamefont {S.}~\bibnamefont {Zhang}},\ and\ \bibinfo {author} {\bibfnamefont {H.~B.}\ \bibnamefont {Sun}},\ }\bibfield  {title} {\bibinfo {title} {Directional forces by momentumless excitation and order-to-order transition in peierls-distorted solids: The case of \ce{GeTe}},\ }\href {https://doi.org/10.1103/PhysRevLett.120.185701} {\bibfield  {journal} {\bibinfo  {journal} {Phys Rev Lett}\ }\textbf {\bibinfo {volume} {120}},\ \bibinfo {pages} {185701} (\bibinfo {year} {2018})}\BibitemShut {NoStop}%
\bibitem [{\citenamefont {Chen}\ \emph {et~al.}(2023)\citenamefont {Chen}, \citenamefont {Wang}, \citenamefont {Jiang}, \citenamefont {Zhang}, \citenamefont {Li}, \citenamefont {Shang}, \citenamefont {Zhu}, \citenamefont {Gong},\ and\ \citenamefont {Hu}}]{Chen:JPCL:2023}%
  \BibitemOpen
  \bibfield  {author} {\bibinfo {author} {\bibfnamefont {L.}~\bibnamefont {Chen}}, \bibinfo {author} {\bibfnamefont {L.}~\bibnamefont {Wang}}, \bibinfo {author} {\bibfnamefont {K.}~\bibnamefont {Jiang}}, \bibinfo {author} {\bibfnamefont {J.}~\bibnamefont {Zhang}}, \bibinfo {author} {\bibfnamefont {Y.}~\bibnamefont {Li}}, \bibinfo {author} {\bibfnamefont {L.}~\bibnamefont {Shang}}, \bibinfo {author} {\bibfnamefont {L.}~\bibnamefont {Zhu}}, \bibinfo {author} {\bibfnamefont {S.}~\bibnamefont {Gong}},\ and\ \bibinfo {author} {\bibfnamefont {Z.}~\bibnamefont {Hu}},\ }\bibfield  {title} {\bibinfo {title} {Optically induced multistage phase transition in coherent phonon-dominated \ce{a-GeTe}},\ }\href {https://doi.org/10.1021/acs.jpclett.3c01173} {\bibfield  {journal} {\bibinfo  {journal} {J Phys Chem Lett}\ }\textbf {\bibinfo {volume} {14}},\ \bibinfo {pages} {5760} (\bibinfo {year} {2023})}\BibitemShut {NoStop}%
\bibitem [{\citenamefont {Miller}\ \emph {et~al.}(2016)\citenamefont {Miller}, \citenamefont {Rudé}, \citenamefont {Pruneri},\ and\ \citenamefont {Wall}}]{Miller:PRB:2016}%
  \BibitemOpen
  \bibfield  {author} {\bibinfo {author} {\bibfnamefont {T.~A.}\ \bibnamefont {Miller}}, \bibinfo {author} {\bibfnamefont {M.}~\bibnamefont {Rudé}}, \bibinfo {author} {\bibfnamefont {V.}~\bibnamefont {Pruneri}},\ and\ \bibinfo {author} {\bibfnamefont {S.}~\bibnamefont {Wall}},\ }\bibfield  {title} {\bibinfo {title} {Ultrafast optical response of the amorphous and crystalline states of the phase change material \ce{Ge2Sb2Te5}},\ }\bibfield  {journal} {\bibinfo  {journal} {Physical Review B}\ }\textbf {\bibinfo {volume} {94}},\ \href {https://doi.org/10.1103/PhysRevB.94.024301} {10.1103/PhysRevB.94.024301} (\bibinfo {year} {2016})\BibitemShut {NoStop}%
\bibitem [{\citenamefont {van Pieterson}\ \emph {et~al.}(2005)\citenamefont {van Pieterson}, \citenamefont {Lankhorst}, \citenamefont {van Schijndel}, \citenamefont {Kuiper},\ and\ \citenamefont {Roosen}}]{Pieterson:JAP:2005}%
  \BibitemOpen
  \bibfield  {author} {\bibinfo {author} {\bibfnamefont {L.}~\bibnamefont {van Pieterson}}, \bibinfo {author} {\bibfnamefont {M.~H.~R.}\ \bibnamefont {Lankhorst}}, \bibinfo {author} {\bibfnamefont {M.}~\bibnamefont {van Schijndel}}, \bibinfo {author} {\bibfnamefont {A.~E.~T.}\ \bibnamefont {Kuiper}},\ and\ \bibinfo {author} {\bibfnamefont {J.~H.~J.}\ \bibnamefont {Roosen}},\ }\bibfield  {title} {\bibinfo {title} {Phase-change recording materials with a growth-dominated crystallization mechanism: A materials overview},\ }\bibfield  {journal} {\bibinfo  {journal} {Journal of Applied Physics}\ }\textbf {\bibinfo {volume} {97}},\ \href {https://doi.org/10.1063/1.1868860} {10.1063/1.1868860} (\bibinfo {year} {2005})\BibitemShut {NoStop}%
\bibitem [{\citenamefont {Siwick}\ \emph {et~al.}(2003)\citenamefont {Siwick}, \citenamefont {Dwyer}, \citenamefont {Jordan},\ and\ \citenamefont {Miller}}]{Siwick:science:2003}%
  \BibitemOpen
  \bibfield  {author} {\bibinfo {author} {\bibfnamefont {B.~J.}\ \bibnamefont {Siwick}}, \bibinfo {author} {\bibfnamefont {J.~R.}\ \bibnamefont {Dwyer}}, \bibinfo {author} {\bibfnamefont {R.~E.}\ \bibnamefont {Jordan}},\ and\ \bibinfo {author} {\bibfnamefont {R.~J.}\ \bibnamefont {Miller}},\ }\bibfield  {title} {\bibinfo {title} {An atomic-level view of melting using femtosecond electron diffraction},\ }\href {https://doi.org/10.1126/science.1090052} {\bibfield  {journal} {\bibinfo  {journal} {Science}\ }\textbf {\bibinfo {volume} {302}},\ \bibinfo {pages} {1382} (\bibinfo {year} {2003})}\BibitemShut {NoStop}%
\bibitem [{\citenamefont {Zong}\ \emph {et~al.}(2021)\citenamefont {Zong}, \citenamefont {Kogar},\ and\ \citenamefont {Gedik}}]{Zong:MRSreview:2021}%
  \BibitemOpen
  \bibfield  {author} {\bibinfo {author} {\bibfnamefont {A.}~\bibnamefont {Zong}}, \bibinfo {author} {\bibfnamefont {A.}~\bibnamefont {Kogar}},\ and\ \bibinfo {author} {\bibfnamefont {N.}~\bibnamefont {Gedik}},\ }\bibfield  {title} {\bibinfo {title} {Unconventional light-induced states visualized by ultrafast electron diffraction and microscopy},\ }\href {https://doi.org/10.1557/s43577-021-00163-8} {\bibfield  {journal} {\bibinfo  {journal} {MRS Bulletin}\ }\textbf {\bibinfo {volume} {46}},\ \bibinfo {pages} {720} (\bibinfo {year} {2021})}\BibitemShut {NoStop}%
\bibitem [{\citenamefont {Shen}(1989)}]{Shen:SFG:1989}%
  \BibitemOpen
  \bibfield  {author} {\bibinfo {author} {\bibfnamefont {Y.~R.}\ \bibnamefont {Shen}},\ }\bibfield  {title} {\bibinfo {title} {Surface properties probed by second-harmonic and sum-frequency generation},\ }\href {https://doi.org/10.1038/337519a0} {\bibfield  {journal} {\bibinfo  {journal} {Nature}\ }\textbf {\bibinfo {volume} {337}},\ \bibinfo {pages} {519} (\bibinfo {year} {1989})}\BibitemShut {NoStop}%
\bibitem [{\citenamefont {Foglia}\ \emph {et~al.}(2016)\citenamefont {Foglia}, \citenamefont {Wolf},\ and\ \citenamefont {Stähler}}]{Foglia:APL:2016}%
  \BibitemOpen
  \bibfield  {author} {\bibinfo {author} {\bibfnamefont {L.}~\bibnamefont {Foglia}}, \bibinfo {author} {\bibfnamefont {M.}~\bibnamefont {Wolf}},\ and\ \bibinfo {author} {\bibfnamefont {J.}~\bibnamefont {Stähler}},\ }\bibfield  {title} {\bibinfo {title} {Ultrafast dynamics in solids probed by femtosecond time-resolved broadband electronic sum frequency generation},\ }\bibfield  {journal} {\bibinfo  {journal} {Applied Physics Letters}\ }\textbf {\bibinfo {volume} {109}},\ \href {https://doi.org/10.1063/1.4967838} {10.1063/1.4967838} (\bibinfo {year} {2016})\BibitemShut {NoStop}%
\bibitem [{\citenamefont {Wang}\ \emph {et~al.}(2018)\citenamefont {Wang}, \citenamefont {Wang}, \citenamefont {Zheng}, \citenamefont {Liu}, \citenamefont {Li}, \citenamefont {Lv}, \citenamefont {Song}, \citenamefont {Song}, \citenamefont {Cheng}, \citenamefont {Ren},\ and\ \citenamefont {Song}}]{Wang:Sb2Te:2018}%
  \BibitemOpen
  \bibfield  {author} {\bibinfo {author} {\bibfnamefont {Y.}~\bibnamefont {Wang}}, \bibinfo {author} {\bibfnamefont {T.}~\bibnamefont {Wang}}, \bibinfo {author} {\bibfnamefont {Y.}~\bibnamefont {Zheng}}, \bibinfo {author} {\bibfnamefont {G.}~\bibnamefont {Liu}}, \bibinfo {author} {\bibfnamefont {T.}~\bibnamefont {Li}}, \bibinfo {author} {\bibfnamefont {S.}~\bibnamefont {Lv}}, \bibinfo {author} {\bibfnamefont {W.}~\bibnamefont {Song}}, \bibinfo {author} {\bibfnamefont {S.}~\bibnamefont {Song}}, \bibinfo {author} {\bibfnamefont {Y.}~\bibnamefont {Cheng}}, \bibinfo {author} {\bibfnamefont {K.}~\bibnamefont {Ren}},\ and\ \bibinfo {author} {\bibfnamefont {Z.}~\bibnamefont {Song}},\ }\bibfield  {title} {\bibinfo {title} {Atomic scale insight into the effects of aluminum doped \ce{Sb2Te} for phase change memory application},\ }\href {https://doi.org/10.1038/s41598-018-33421-y} {\bibfield  {journal} {\bibinfo  {journal} {Sci Rep}\ }\textbf {\bibinfo {volume} {8}},\ \bibinfo {pages} {15136} (\bibinfo {year} {2018})}\BibitemShut {NoStop}%
\bibitem [{\citenamefont {Zhu}\ \emph {et~al.}(2019)\citenamefont {Zhu}, \citenamefont {Song}, \citenamefont {Konze}, \citenamefont {Li}, \citenamefont {Gault}, \citenamefont {Chen}, \citenamefont {Shen}, \citenamefont {Lv}, \citenamefont {Song}, \citenamefont {Wuttig},\ and\ \citenamefont {Dronskowski}}]{Zhu:Sb2Te:2019}%
  \BibitemOpen
  \bibfield  {author} {\bibinfo {author} {\bibfnamefont {M.}~\bibnamefont {Zhu}}, \bibinfo {author} {\bibfnamefont {W.}~\bibnamefont {Song}}, \bibinfo {author} {\bibfnamefont {P.~M.}\ \bibnamefont {Konze}}, \bibinfo {author} {\bibfnamefont {T.}~\bibnamefont {Li}}, \bibinfo {author} {\bibfnamefont {B.}~\bibnamefont {Gault}}, \bibinfo {author} {\bibfnamefont {X.}~\bibnamefont {Chen}}, \bibinfo {author} {\bibfnamefont {J.}~\bibnamefont {Shen}}, \bibinfo {author} {\bibfnamefont {S.}~\bibnamefont {Lv}}, \bibinfo {author} {\bibfnamefont {Z.}~\bibnamefont {Song}}, \bibinfo {author} {\bibfnamefont {M.}~\bibnamefont {Wuttig}},\ and\ \bibinfo {author} {\bibfnamefont {R.}~\bibnamefont {Dronskowski}},\ }\bibfield  {title} {\bibinfo {title} {Direct atomic insight into the role of dopants in phase-change materials},\ }\href {https://doi.org/10.1038/s41467-019-11506-0} {\bibfield  {journal} {\bibinfo  {journal} {Nat Commun}\ }\textbf {\bibinfo {volume} {10}},\ \bibinfo {pages} {3525} (\bibinfo {year} {2019})}\BibitemShut {NoStop}%
\bibitem [{\citenamefont {Coppens}(2006)}]{Coppens:ITC:2006}%
  \BibitemOpen
  \bibfield  {author} {\bibinfo {author} {\bibfnamefont {P.}~\bibnamefont {Coppens}},\ }\bibinfo {title} {The structure factor},\ in\ \href {https://doi.org/10.1107/97809553602060000550} {\emph {\bibinfo {booktitle} {International Tables for Crystallography}}},\ \bibinfo {series and number} {International Tables for Crystallography}\ (\bibinfo {year} {2006})\ \bibinfo {type} {Book section}\ \bibinfo {chapter} {Chapter 1.2}, pp.\ \bibinfo {pages} {10--24}\BibitemShut {NoStop}%
\bibitem [{\citenamefont {Huang}\ \emph {et~al.}(2024)\citenamefont {Huang}, \citenamefont {Roos}, \citenamefont {Tong},\ and\ \citenamefont {Campen}}]{Huang:RSI:2024}%
  \BibitemOpen
  \bibfield  {author} {\bibinfo {author} {\bibfnamefont {Z.}~\bibnamefont {Huang}}, \bibinfo {author} {\bibfnamefont {T.}~\bibnamefont {Roos}}, \bibinfo {author} {\bibfnamefont {Y.}~\bibnamefont {Tong}},\ and\ \bibinfo {author} {\bibfnamefont {R.~K.}\ \bibnamefont {Campen}},\ }\bibfield  {title} {\bibinfo {title} {Integration of conventional surface science techniques with surface-sensitive azimuthal and polarization dependent femtosecond-resolved sum frequency generation spectroscopy},\ }\bibfield  {journal} {\bibinfo  {journal} {Rev. Sci. Inst.}\ }\textbf {\bibinfo {volume} {95}},\ \href {https://doi.org/10.1063/5.0205278} {10.1063/5.0205278} (\bibinfo {year} {2024})\BibitemShut {NoStop}%
\bibitem [{\citenamefont {Ishikawa}\ \emph {et~al.}(2015)\citenamefont {Ishikawa}, \citenamefont {Hayes}, \citenamefont {Keskin}, \citenamefont {Corthey}, \citenamefont {Hada}, \citenamefont {Pichugin}, \citenamefont {Marx}, \citenamefont {Hirscht}, \citenamefont {Shionuma}, \citenamefont {Onda}, \citenamefont {Okimoto}, \citenamefont {Koshihara}, \citenamefont {Yamamoto}, \citenamefont {Cui}, \citenamefont {Nomura}, \citenamefont {Oshima}, \citenamefont {Abdel-Jawad}, \citenamefont {Kato},\ and\ \citenamefont {Miller}}]{Ishikawa:science:2015}%
  \BibitemOpen
  \bibfield  {author} {\bibinfo {author} {\bibfnamefont {T.}~\bibnamefont {Ishikawa}}, \bibinfo {author} {\bibfnamefont {S.~A.}\ \bibnamefont {Hayes}}, \bibinfo {author} {\bibfnamefont {S.}~\bibnamefont {Keskin}}, \bibinfo {author} {\bibfnamefont {G.}~\bibnamefont {Corthey}}, \bibinfo {author} {\bibfnamefont {M.}~\bibnamefont {Hada}}, \bibinfo {author} {\bibfnamefont {K.}~\bibnamefont {Pichugin}}, \bibinfo {author} {\bibfnamefont {A.}~\bibnamefont {Marx}}, \bibinfo {author} {\bibfnamefont {J.}~\bibnamefont {Hirscht}}, \bibinfo {author} {\bibfnamefont {K.}~\bibnamefont {Shionuma}}, \bibinfo {author} {\bibfnamefont {K.}~\bibnamefont {Onda}}, \bibinfo {author} {\bibfnamefont {Y.}~\bibnamefont {Okimoto}}, \bibinfo {author} {\bibfnamefont {S.~Y.}\ \bibnamefont {Koshihara}}, \bibinfo {author} {\bibfnamefont {T.}~\bibnamefont {Yamamoto}}, \bibinfo {author} {\bibfnamefont {H.}~\bibnamefont {Cui}}, \bibinfo {author} {\bibfnamefont {M.}~\bibnamefont {Nomura}}, \bibinfo {author} {\bibfnamefont {Y.}~\bibnamefont {Oshima}}, \bibinfo {author} {\bibfnamefont {M.}~\bibnamefont {Abdel-Jawad}}, \bibinfo {author} {\bibfnamefont {R.}~\bibnamefont {Kato}},\ and\ \bibinfo {author} {\bibfnamefont {R.~J.}\ \bibnamefont {Miller}},\ }\bibfield  {title} {\bibinfo {title} {Direct observation of collective modes coupled to molecular orbital-driven charge transfer},\ }\href {https://doi.org/10.1126/science.aab3480} {\bibfield  {journal} {\bibinfo  {journal} {Science}\ }\textbf {\bibinfo {volume} {350}},\ \bibinfo {pages} {1501} (\bibinfo {year} {2015})}\BibitemShut {NoStop}%
\bibitem [{\citenamefont {Zahn}\ \emph {et~al.}(2021)\citenamefont {Zahn}, \citenamefont {Seiler}, \citenamefont {Windsor},\ and\ \citenamefont {Ernstorfer}}]{Zahn:SD:2021}%
  \BibitemOpen
  \bibfield  {author} {\bibinfo {author} {\bibfnamefont {D.}~\bibnamefont {Zahn}}, \bibinfo {author} {\bibfnamefont {H.}~\bibnamefont {Seiler}}, \bibinfo {author} {\bibfnamefont {Y.~W.}\ \bibnamefont {Windsor}},\ and\ \bibinfo {author} {\bibfnamefont {R.}~\bibnamefont {Ernstorfer}},\ }\bibfield  {title} {\bibinfo {title} {Ultrafast lattice dynamics and electron-phonon coupling in platinum extracted with a global fitting approach for time-resolved polycrystalline diffraction data},\ }\href {https://doi.org/10.1063/4.0000120} {\bibfield  {journal} {\bibinfo  {journal} {Struct Dyn}\ }\textbf {\bibinfo {volume} {8}},\ \bibinfo {pages} {064301} (\bibinfo {year} {2021})}\BibitemShut {NoStop}%
\bibitem [{\citenamefont {Zheng}\ \emph {et~al.}(2016)\citenamefont {Zheng}, \citenamefont {Xia}, \citenamefont {Cheng}, \citenamefont {Rao}, \citenamefont {Ding}, \citenamefont {Liu}, \citenamefont {Jia}, \citenamefont {Song},\ and\ \citenamefont {Feng}}]{Zheng:nanores:2016}%
  \BibitemOpen
  \bibfield  {author} {\bibinfo {author} {\bibfnamefont {Y.~H.}\ \bibnamefont {Zheng}}, \bibinfo {author} {\bibfnamefont {M.~J.}\ \bibnamefont {Xia}}, \bibinfo {author} {\bibfnamefont {Y.}~\bibnamefont {Cheng}}, \bibinfo {author} {\bibfnamefont {F.}~\bibnamefont {Rao}}, \bibinfo {author} {\bibfnamefont {K.~Y.}\ \bibnamefont {Ding}}, \bibinfo {author} {\bibfnamefont {W.~L.}\ \bibnamefont {Liu}}, \bibinfo {author} {\bibfnamefont {Y.}~\bibnamefont {Jia}}, \bibinfo {author} {\bibfnamefont {Z.~T.}\ \bibnamefont {Song}},\ and\ \bibinfo {author} {\bibfnamefont {S.~L.}\ \bibnamefont {Feng}},\ }\bibfield  {title} {\bibinfo {title} {Direct observation of metastable face-centered cubic \ce{Sb2Te3} crystal},\ }\href {https://doi.org/10.1007/s12274-016-1221-8} {\bibfield  {journal} {\bibinfo  {journal} {Nano Research}\ }\textbf {\bibinfo {volume} {9}},\ \bibinfo {pages} {3453} (\bibinfo {year} {2016})}\BibitemShut {NoStop}%
\bibitem [{\citenamefont {Peng}\ \emph {et~al.}(2004)\citenamefont {Peng}, \citenamefont {Dudarev},\ and\ \citenamefont {Whelan}}]{Peng:DW:2004}%
  \BibitemOpen
  \bibfield  {author} {\bibinfo {author} {\bibfnamefont {L.~M.}\ \bibnamefont {Peng}}, \bibinfo {author} {\bibfnamefont {S.~L.}\ \bibnamefont {Dudarev}},\ and\ \bibinfo {author} {\bibfnamefont {M.~J.}\ \bibnamefont {Whelan}},\ }\href {https://cir.nii.ac.jp/crid/1130000795485621120} {\emph {\bibinfo {title} {High-energy electron diffraction and microscopy}}},\ Monographs on the physics and chemistry of materials\ (\bibinfo  {publisher} {Oxford University Press},\ \bibinfo {year} {2004})\BibitemShut {NoStop}%
\bibitem [{\citenamefont {Stern}\ \emph {et~al.}(2018)\citenamefont {Stern}, \citenamefont {René~de Cotret}, \citenamefont {Otto}, \citenamefont {Chatelain}, \citenamefont {Boisvert}, \citenamefont {Sutton},\ and\ \citenamefont {Siwick}}]{Stern:PRB:2018}%
  \BibitemOpen
  \bibfield  {author} {\bibinfo {author} {\bibfnamefont {M.~J.}\ \bibnamefont {Stern}}, \bibinfo {author} {\bibfnamefont {L.~P.}\ \bibnamefont {René~de Cotret}}, \bibinfo {author} {\bibfnamefont {M.~R.}\ \bibnamefont {Otto}}, \bibinfo {author} {\bibfnamefont {R.~P.}\ \bibnamefont {Chatelain}}, \bibinfo {author} {\bibfnamefont {J.-P.}\ \bibnamefont {Boisvert}}, \bibinfo {author} {\bibfnamefont {M.}~\bibnamefont {Sutton}},\ and\ \bibinfo {author} {\bibfnamefont {B.~J.}\ \bibnamefont {Siwick}},\ }\bibfield  {title} {\bibinfo {title} {Mapping momentum-dependent electron-phonon coupling and nonequilibrium phonon dynamics with ultrafast electron diffuse scattering},\ }\bibfield  {journal} {\bibinfo  {journal} {Phys. Rev. B}\ }\textbf {\bibinfo {volume} {97}},\ \href {https://doi.org/10.1103/PhysRevB.97.165416} {10.1103/PhysRevB.97.165416} (\bibinfo {year} {2018})\BibitemShut {NoStop}%
\bibitem [{\citenamefont {Durr}\ \emph {et~al.}(2021)\citenamefont {Durr}, \citenamefont {Ernstorfer},\ and\ \citenamefont {Siwick}}]{Durr:MRS:2021}%
  \BibitemOpen
  \bibfield  {author} {\bibinfo {author} {\bibfnamefont {H.~A.}\ \bibnamefont {Durr}}, \bibinfo {author} {\bibfnamefont {R.}~\bibnamefont {Ernstorfer}},\ and\ \bibinfo {author} {\bibfnamefont {B.~J.}\ \bibnamefont {Siwick}},\ }\bibfield  {title} {\bibinfo {title} {Revealing momentum-dependent electron-phonon and phonon-phonon coupling in complex materials with ultrafast electron diffuse scattering},\ }\href {https://doi.org/10.1557/s43577-021-00156-7} {\bibfield  {journal} {\bibinfo  {journal} {MRS Bull.}\ }\textbf {\bibinfo {volume} {46}},\ \bibinfo {pages} {731} (\bibinfo {year} {2021})}\BibitemShut {NoStop}%
\bibitem [{\citenamefont {Gaspard}(2016)}]{Gaspard:Peierls:2016}%
  \BibitemOpen
  \bibfield  {author} {\bibinfo {author} {\bibfnamefont {J.-P.}\ \bibnamefont {Gaspard}},\ }\bibfield  {title} {\bibinfo {title} {Structure of covalently bonded materials: From the peierls distortion to phase-change materials},\ }\href {https://doi.org/10.1016/j.crhy.2015.12.009} {\bibfield  {journal} {\bibinfo  {journal} {Comptes Rendus Physique}\ }\textbf {\bibinfo {volume} {17}},\ \bibinfo {pages} {389} (\bibinfo {year} {2016})}\BibitemShut {NoStop}%
\bibitem [{\citenamefont {Wall}\ \emph {et~al.}(2018)\citenamefont {Wall}, \citenamefont {Yang}, \citenamefont {Vidas}, \citenamefont {Chollet}, \citenamefont {Glownia}, \citenamefont {Kozina}, \citenamefont {Katayama}, \citenamefont {Henighan}, \citenamefont {Jiang}, \citenamefont {Miller}, \citenamefont {Reis}, \citenamefont {Boatner}, \citenamefont {Delaire},\ and\ \citenamefont {Trigo}}]{Wall:VO2:2018}%
  \BibitemOpen
  \bibfield  {author} {\bibinfo {author} {\bibfnamefont {S.}~\bibnamefont {Wall}}, \bibinfo {author} {\bibfnamefont {S.}~\bibnamefont {Yang}}, \bibinfo {author} {\bibfnamefont {L.}~\bibnamefont {Vidas}}, \bibinfo {author} {\bibfnamefont {M.}~\bibnamefont {Chollet}}, \bibinfo {author} {\bibfnamefont {J.~M.}\ \bibnamefont {Glownia}}, \bibinfo {author} {\bibfnamefont {M.}~\bibnamefont {Kozina}}, \bibinfo {author} {\bibfnamefont {T.}~\bibnamefont {Katayama}}, \bibinfo {author} {\bibfnamefont {T.}~\bibnamefont {Henighan}}, \bibinfo {author} {\bibfnamefont {M.}~\bibnamefont {Jiang}}, \bibinfo {author} {\bibfnamefont {T.~A.}\ \bibnamefont {Miller}}, \bibinfo {author} {\bibfnamefont {D.~A.}\ \bibnamefont {Reis}}, \bibinfo {author} {\bibfnamefont {L.~A.}\ \bibnamefont {Boatner}}, \bibinfo {author} {\bibfnamefont {O.}~\bibnamefont {Delaire}},\ and\ \bibinfo {author} {\bibfnamefont {M.}~\bibnamefont {Trigo}},\ }\bibfield  {title} {\bibinfo {title} {Ultrafast disordering of vanadium dimers in photoexcited \ce{VO2}},\ }\href {https://doi.org/10.1126/science.aau3873} {\bibfield  {journal} {\bibinfo  {journal} {Science}\ }\textbf {\bibinfo {volume} {362}},\ \bibinfo {pages} {572} (\bibinfo {year} {2018})}\BibitemShut {NoStop}%
\bibitem [{\citenamefont {de~la Peña~Muñoz}\ \emph {et~al.}(2023)\citenamefont {de~la Peña~Muñoz}, \citenamefont {Correa}, \citenamefont {Yang}, \citenamefont {Delaire}, \citenamefont {Huang}, \citenamefont {Johnson}, \citenamefont {Katayama}, \citenamefont {Krapivin}, \citenamefont {Pastor}, \citenamefont {Reis}, \citenamefont {Teitelbaum}, \citenamefont {Vidas}, \citenamefont {Wall},\ and\ \citenamefont {Trigo}}]{De:NP:2023}%
  \BibitemOpen
  \bibfield  {author} {\bibinfo {author} {\bibfnamefont {G.~A.}\ \bibnamefont {de~la Peña~Muñoz}}, \bibinfo {author} {\bibfnamefont {A.~A.}\ \bibnamefont {Correa}}, \bibinfo {author} {\bibfnamefont {S.}~\bibnamefont {Yang}}, \bibinfo {author} {\bibfnamefont {O.}~\bibnamefont {Delaire}}, \bibinfo {author} {\bibfnamefont {Y.}~\bibnamefont {Huang}}, \bibinfo {author} {\bibfnamefont {A.~S.}\ \bibnamefont {Johnson}}, \bibinfo {author} {\bibfnamefont {T.}~\bibnamefont {Katayama}}, \bibinfo {author} {\bibfnamefont {V.}~\bibnamefont {Krapivin}}, \bibinfo {author} {\bibfnamefont {E.}~\bibnamefont {Pastor}}, \bibinfo {author} {\bibfnamefont {D.~A.}\ \bibnamefont {Reis}}, \bibinfo {author} {\bibfnamefont {S.}~\bibnamefont {Teitelbaum}}, \bibinfo {author} {\bibfnamefont {L.}~\bibnamefont {Vidas}}, \bibinfo {author} {\bibfnamefont {S.}~\bibnamefont {Wall}},\ and\ \bibinfo {author} {\bibfnamefont {M.}~\bibnamefont {Trigo}},\ }\bibfield  {title} {\bibinfo {title} {Ultrafast lattice disordering can be accelerated by electronic collisional forces},\ }\href {https://doi.org/10.1038/s41567-023-02118-z} {\bibfield  {journal} {\bibinfo  {journal} {Nature Physics}\ }\textbf {\bibinfo {volume} {19}},\ \bibinfo {pages} {1489} (\bibinfo {year} {2023})}\BibitemShut {NoStop}%
\bibitem [{\citenamefont {Hohlfeld}\ \emph {et~al.}(2000)\citenamefont {Hohlfeld}, \citenamefont {Wellershoff}, \citenamefont {Güdde}, \citenamefont {Conrad}, \citenamefont {Jähnke},\ and\ \citenamefont {Matthias}}]{Hohlfeld:CP:2000}%
  \BibitemOpen
  \bibfield  {author} {\bibinfo {author} {\bibfnamefont {J.}~\bibnamefont {Hohlfeld}}, \bibinfo {author} {\bibfnamefont {S.~S.}\ \bibnamefont {Wellershoff}}, \bibinfo {author} {\bibfnamefont {J.}~\bibnamefont {Güdde}}, \bibinfo {author} {\bibfnamefont {U.}~\bibnamefont {Conrad}}, \bibinfo {author} {\bibfnamefont {V.}~\bibnamefont {Jähnke}},\ and\ \bibinfo {author} {\bibfnamefont {E.}~\bibnamefont {Matthias}},\ }\bibfield  {title} {\bibinfo {title} {Electron and lattice dynamics following optical excitation of metals},\ }\href {https://doi.org/10.1016/s0301-0104(99)00330-4} {\bibfield  {journal} {\bibinfo  {journal} {Chemical Physics}\ }\textbf {\bibinfo {volume} {251}},\ \bibinfo {pages} {237} (\bibinfo {year} {2000})}\BibitemShut {NoStop}%
\bibitem [{\citenamefont {Agafonov}\ \emph {et~al.}(1991)\citenamefont {Agafonov}, \citenamefont {Rodier}, \citenamefont {Ceolin}, \citenamefont {Bellissent}, \citenamefont {Bergman},\ and\ \citenamefont {Gaspard}}]{Agafonov:Sb2Te:1991}%
  \BibitemOpen
  \bibfield  {author} {\bibinfo {author} {\bibfnamefont {V.}~\bibnamefont {Agafonov}}, \bibinfo {author} {\bibfnamefont {N.}~\bibnamefont {Rodier}}, \bibinfo {author} {\bibfnamefont {R.}~\bibnamefont {Ceolin}}, \bibinfo {author} {\bibfnamefont {R.}~\bibnamefont {Bellissent}}, \bibinfo {author} {\bibfnamefont {C.}~\bibnamefont {Bergman}},\ and\ \bibinfo {author} {\bibfnamefont {J.~P.}\ \bibnamefont {Gaspard}},\ }\bibfield  {title} {\bibinfo {title} {Structure of sb$_2$te},\ }\href {https://doi.org/Doi 10.1107/S0108270190013348} {\bibfield  {journal} {\bibinfo  {journal} {Acta Crystallographica Section C-Crystal Structure Communications}\ }\textbf {\bibinfo {volume} {47}},\ \bibinfo {pages} {1141} (\bibinfo {year} {1991})}\BibitemShut {NoStop}%
\bibitem [{\citenamefont {Sokolowski-Tinten}\ \emph {et~al.}(2003)\citenamefont {Sokolowski-Tinten}, \citenamefont {Blome}, \citenamefont {Blums}, \citenamefont {Cavalleri}, \citenamefont {Dietrich}, \citenamefont {Tarasevitch}, \citenamefont {Uschmann}, \citenamefont {Forster}, \citenamefont {Kammler}, \citenamefont {Horn-von Hoegen},\ and\ \citenamefont {von~der Linde}}]{Sokolowski-Tinten:nature:2003}%
  \BibitemOpen
  \bibfield  {author} {\bibinfo {author} {\bibfnamefont {K.}~\bibnamefont {Sokolowski-Tinten}}, \bibinfo {author} {\bibfnamefont {C.}~\bibnamefont {Blome}}, \bibinfo {author} {\bibfnamefont {J.}~\bibnamefont {Blums}}, \bibinfo {author} {\bibfnamefont {A.}~\bibnamefont {Cavalleri}}, \bibinfo {author} {\bibfnamefont {C.}~\bibnamefont {Dietrich}}, \bibinfo {author} {\bibfnamefont {A.}~\bibnamefont {Tarasevitch}}, \bibinfo {author} {\bibfnamefont {I.}~\bibnamefont {Uschmann}}, \bibinfo {author} {\bibfnamefont {E.}~\bibnamefont {Forster}}, \bibinfo {author} {\bibfnamefont {M.}~\bibnamefont {Kammler}}, \bibinfo {author} {\bibfnamefont {M.}~\bibnamefont {Horn-von Hoegen}},\ and\ \bibinfo {author} {\bibfnamefont {D.}~\bibnamefont {von~der Linde}},\ }\bibfield  {title} {\bibinfo {title} {Femtosecond x-ray measurement of coherent lattice vibrations near the lindemann stability limit},\ }\href {https://doi.org/10.1038/nature01490} {\bibfield  {journal} {\bibinfo  {journal} {Nature}\ }\textbf {\bibinfo {volume} {422}},\ \bibinfo {pages} {287} (\bibinfo {year} {2003})}\BibitemShut {NoStop}%
\bibitem [{\citenamefont {Fritz}\ \emph {et~al.}(2007)\citenamefont {Fritz}, \citenamefont {Reis}, \citenamefont {Adams}, \citenamefont {Akre}, \citenamefont {Arthur}, \citenamefont {Blome}, \citenamefont {Bucksbaum}, \citenamefont {Cavalieri}, \citenamefont {Engemann}, \citenamefont {Fahy}, \citenamefont {Falcone}, \citenamefont {Fuoss}, \citenamefont {Gaffney}, \citenamefont {George}, \citenamefont {Hajdu}, \citenamefont {Hertlein}, \citenamefont {Hillyard}, \citenamefont {Horn-von Hoegen}, \citenamefont {Kammler}, \citenamefont {Kaspar}, \citenamefont {Kienberger}, \citenamefont {Krejcik}, \citenamefont {Lee}, \citenamefont {Lindenberg}, \citenamefont {McFarland}, \citenamefont {Meyer}, \citenamefont {Montagne}, \citenamefont {Murray}, \citenamefont {Nelson}, \citenamefont {Nicoul}, \citenamefont {Pahl}, \citenamefont {Rudati}, \citenamefont {Schlarb}, \citenamefont {Siddons}, \citenamefont {Sokolowski-Tinten}, \citenamefont {Tschentscher}, \citenamefont {von~der Linde},\ and\ \citenamefont {Hastings}}]{Fritz:science:2007}%
  \BibitemOpen
  \bibfield  {author} {\bibinfo {author} {\bibfnamefont {D.~M.}\ \bibnamefont {Fritz}}, \bibinfo {author} {\bibfnamefont {D.~A.}\ \bibnamefont {Reis}}, \bibinfo {author} {\bibfnamefont {B.}~\bibnamefont {Adams}}, \bibinfo {author} {\bibfnamefont {R.~A.}\ \bibnamefont {Akre}}, \bibinfo {author} {\bibfnamefont {J.}~\bibnamefont {Arthur}}, \bibinfo {author} {\bibfnamefont {C.}~\bibnamefont {Blome}}, \bibinfo {author} {\bibfnamefont {P.~H.}\ \bibnamefont {Bucksbaum}}, \bibinfo {author} {\bibfnamefont {A.~L.}\ \bibnamefont {Cavalieri}}, \bibinfo {author} {\bibfnamefont {S.}~\bibnamefont {Engemann}}, \bibinfo {author} {\bibfnamefont {S.}~\bibnamefont {Fahy}}, \bibinfo {author} {\bibfnamefont {R.~W.}\ \bibnamefont {Falcone}}, \bibinfo {author} {\bibfnamefont {P.~H.}\ \bibnamefont {Fuoss}}, \bibinfo {author} {\bibfnamefont {K.~J.}\ \bibnamefont {Gaffney}}, \bibinfo {author} {\bibfnamefont {M.~J.}\ \bibnamefont {George}}, \bibinfo {author} {\bibfnamefont {J.}~\bibnamefont {Hajdu}}, \bibinfo {author} {\bibfnamefont {M.~P.}\ \bibnamefont {Hertlein}}, \bibinfo {author} {\bibfnamefont {P.~B.}\ \bibnamefont {Hillyard}}, \bibinfo {author} {\bibfnamefont {M.}~\bibnamefont {Horn-von Hoegen}}, \bibinfo {author} {\bibfnamefont {M.}~\bibnamefont {Kammler}}, \bibinfo {author} {\bibfnamefont {J.}~\bibnamefont {Kaspar}}, \bibinfo {author} {\bibfnamefont {R.}~\bibnamefont {Kienberger}}, \bibinfo {author} {\bibfnamefont {P.}~\bibnamefont {Krejcik}}, \bibinfo {author} {\bibfnamefont {S.~H.}\ \bibnamefont {Lee}}, \bibinfo {author} {\bibfnamefont {A.~M.}\ \bibnamefont {Lindenberg}}, \bibinfo {author} {\bibfnamefont {B.}~\bibnamefont {McFarland}}, \bibinfo {author} {\bibfnamefont {D.}~\bibnamefont {Meyer}}, \bibinfo {author} {\bibfnamefont {T.}~\bibnamefont {Montagne}}, \bibinfo {author} {\bibfnamefont {E.~D.}\ \bibnamefont {Murray}}, \bibinfo {author} {\bibfnamefont {A.~J.}\ \bibnamefont {Nelson}}, \bibinfo {author} {\bibfnamefont {M.}~\bibnamefont {Nicoul}}, \bibinfo {author} {\bibfnamefont {R.}~\bibnamefont {Pahl}}, \bibinfo {author} {\bibfnamefont {J.}~\bibnamefont {Rudati}}, \bibinfo {author} {\bibfnamefont {H.}~\bibnamefont {Schlarb}}, \bibinfo {author} {\bibfnamefont {D.~P.}\ \bibnamefont {Siddons}}, \bibinfo {author} {\bibfnamefont {K.}~\bibnamefont {Sokolowski-Tinten}}, \bibinfo {author} {\bibfnamefont {T.}~\bibnamefont {Tschentscher}}, \bibinfo {author} {\bibfnamefont {D.}~\bibnamefont {von~der Linde}},\ and\ \bibinfo {author} {\bibfnamefont {J.~B.}\ \bibnamefont {Hastings}},\ }\bibfield  {title} {\bibinfo {title} {Ultrafast bond softening in bismuth: mapping a solid's interatomic potential with x-rays},\ }\href {https://doi.org/10.1126/science.1135009} {\bibfield  {journal} {\bibinfo  {journal} {Science}\ }\textbf {\bibinfo {volume} {315}},\ \bibinfo {pages} {633} (\bibinfo {year} {2007})}\BibitemShut {NoStop}%
\bibitem [{\citenamefont {Sciaini}\ \emph {et~al.}(2009)\citenamefont {Sciaini}, \citenamefont {Harb}, \citenamefont {Kruglik}, \citenamefont {Payer}, \citenamefont {Hebeisen}, \citenamefont {zu~Heringdorf}, \citenamefont {Yamaguchi}, \citenamefont {Horn-von Hoegen}, \citenamefont {Ernstorfer},\ and\ \citenamefont {Miller}}]{Sciaini:nature:2009}%
  \BibitemOpen
  \bibfield  {author} {\bibinfo {author} {\bibfnamefont {G.}~\bibnamefont {Sciaini}}, \bibinfo {author} {\bibfnamefont {M.}~\bibnamefont {Harb}}, \bibinfo {author} {\bibfnamefont {S.~G.}\ \bibnamefont {Kruglik}}, \bibinfo {author} {\bibfnamefont {T.}~\bibnamefont {Payer}}, \bibinfo {author} {\bibfnamefont {C.~T.}\ \bibnamefont {Hebeisen}}, \bibinfo {author} {\bibfnamefont {F.~J.}\ \bibnamefont {zu~Heringdorf}}, \bibinfo {author} {\bibfnamefont {M.}~\bibnamefont {Yamaguchi}}, \bibinfo {author} {\bibfnamefont {M.}~\bibnamefont {Horn-von Hoegen}}, \bibinfo {author} {\bibfnamefont {R.}~\bibnamefont {Ernstorfer}},\ and\ \bibinfo {author} {\bibfnamefont {R.~J.}\ \bibnamefont {Miller}},\ }\bibfield  {title} {\bibinfo {title} {Electronic acceleration of atomic motions and disordering in bismuth},\ }\href {https://doi.org/10.1038/nature07788} {\bibfield  {journal} {\bibinfo  {journal} {Nature}\ }\textbf {\bibinfo {volume} {458}},\ \bibinfo {pages} {56} (\bibinfo {year} {2009})}\BibitemShut {NoStop}%
\bibitem [{\citenamefont {Kimber}\ \emph {et~al.}(2023)\citenamefont {Kimber}, \citenamefont {Zhang}, \citenamefont {Liang}, \citenamefont {Guzmán-Verri}, \citenamefont {Littlewood}, \citenamefont {Cheng}, \citenamefont {Abernathy}, \citenamefont {Hudspeth}, \citenamefont {Luo}, \citenamefont {Kanatzidis}, \citenamefont {Chatterji}, \citenamefont {Ramirez-Cuesta},\ and\ \citenamefont {Billinge}}]{Kimber:NM:2023}%
  \BibitemOpen
  \bibfield  {author} {\bibinfo {author} {\bibfnamefont {S.~A.~J.}\ \bibnamefont {Kimber}}, \bibinfo {author} {\bibfnamefont {J.}~\bibnamefont {Zhang}}, \bibinfo {author} {\bibfnamefont {C.~H.}\ \bibnamefont {Liang}}, \bibinfo {author} {\bibfnamefont {G.~G.}\ \bibnamefont {Guzmán-Verri}}, \bibinfo {author} {\bibfnamefont {P.~B.}\ \bibnamefont {Littlewood}}, \bibinfo {author} {\bibfnamefont {Y.}~\bibnamefont {Cheng}}, \bibinfo {author} {\bibfnamefont {D.~L.}\ \bibnamefont {Abernathy}}, \bibinfo {author} {\bibfnamefont {J.~M.}\ \bibnamefont {Hudspeth}}, \bibinfo {author} {\bibfnamefont {Z.-Z.}\ \bibnamefont {Luo}}, \bibinfo {author} {\bibfnamefont {M.~G.}\ \bibnamefont {Kanatzidis}}, \bibinfo {author} {\bibfnamefont {T.}~\bibnamefont {Chatterji}}, \bibinfo {author} {\bibfnamefont {A.~J.}\ \bibnamefont {Ramirez-Cuesta}},\ and\ \bibinfo {author} {\bibfnamefont {S.~J.~L.}\ \bibnamefont {Billinge}},\ }\bibfield  {title} {\bibinfo {title} {Dynamic crystallography reveals spontaneous anisotropy in cubic \ce{GeTe}},\ }\bibfield  {journal} {\bibinfo  {journal} {Nature Materials}\ }\href {https://doi.org/10.1038/s41563-023-01483-7} {10.1038/s41563-023-01483-7} (\bibinfo {year} {2023})\BibitemShut {NoStop}%
\bibitem [{\citenamefont {Sun}\ \emph {et~al.}(2012)\citenamefont {Sun}, \citenamefont {Zhou}, \citenamefont {Mao},\ and\ \citenamefont {Ahuja}}]{Sun:PNAS:2012}%
  \BibitemOpen
  \bibfield  {author} {\bibinfo {author} {\bibfnamefont {Z.}~\bibnamefont {Sun}}, \bibinfo {author} {\bibfnamefont {J.}~\bibnamefont {Zhou}}, \bibinfo {author} {\bibfnamefont {H.~K.}\ \bibnamefont {Mao}},\ and\ \bibinfo {author} {\bibfnamefont {R.}~\bibnamefont {Ahuja}},\ }\bibfield  {title} {\bibinfo {title} {Peierls distortion mediated reversible phase transition in \ce{GeTe} under pressure},\ }\href {https://doi.org/10.1073/pnas.1202875109} {\bibfield  {journal} {\bibinfo  {journal} {Proc Natl Acad Sci U S A}\ }\textbf {\bibinfo {volume} {109}},\ \bibinfo {pages} {5948} (\bibinfo {year} {2012})}\BibitemShut {NoStop}%
\bibitem [{\citenamefont {Neu}\ and\ \citenamefont {Schmuttenmaer}(2018)}]{Neu:THz:2018}%
  \BibitemOpen
  \bibfield  {author} {\bibinfo {author} {\bibfnamefont {J.}~\bibnamefont {Neu}}\ and\ \bibinfo {author} {\bibfnamefont {C.~A.}\ \bibnamefont {Schmuttenmaer}},\ }\bibfield  {title} {\bibinfo {title} {Tutorial: An introduction to terahertz time domain spectroscopy (\ce{THz-TDS})},\ }\bibfield  {journal} {\bibinfo  {journal} {J. Appl. Phys.}\ }\textbf {\bibinfo {volume} {124}},\ \href {https://doi.org/10.1063/1.5047659} {10.1063/1.5047659} (\bibinfo {year} {2018})\BibitemShut {NoStop}%
\bibitem [{\citenamefont {Sobota}\ \emph {et~al.}(2021)\citenamefont {Sobota}, \citenamefont {He},\ and\ \citenamefont {Shen}}]{Sobota:arpes:2021}%
  \BibitemOpen
  \bibfield  {author} {\bibinfo {author} {\bibfnamefont {J.~A.}\ \bibnamefont {Sobota}}, \bibinfo {author} {\bibfnamefont {Y.}~\bibnamefont {He}},\ and\ \bibinfo {author} {\bibfnamefont {Z.-X.}\ \bibnamefont {Shen}},\ }\bibfield  {title} {\bibinfo {title} {Angle-resolved photoemission studies of quantum materials},\ }\bibfield  {journal} {\bibinfo  {journal} {Reviews of Modern Physics}\ }\textbf {\bibinfo {volume} {93}},\ \href {https://doi.org/10.1103/RevModPhys.93.025006} {10.1103/RevModPhys.93.025006} (\bibinfo {year} {2021})\BibitemShut {NoStop}%
\end{thebibliography}%

\end{document}